\documentclass{article}
\usepackage[left=3cm,right=3cm,top=3cm,bottom=3cm]{geometry} 
\usepackage{amsmath} 

\usepackage{graphicx}
\usepackage{authblk}

\usepackage{amsfonts}
\usepackage{booktabs}
\usepackage{tabularx}
\usepackage{ltablex}
\usepackage{longtable}

\usepackage{natbib,upgreek}
\usepackage{graphicx,hyperref}
\usepackage{amsmath}
\usepackage{multirow}
\usepackage{float}
\usepackage{mathtools}
\mathtoolsset{showonlyrefs=true}
\usepackage{subfig}
\usepackage{graphicx}
\usepackage{epsfig}
\usepackage{natbib}
\usepackage{color}
\usepackage{xcolor}

\usepackage{fancyhdr,lipsum}
\makeatletter

\newsavebox{\@linebox}
\savebox{\@linebox}[3em][t]{\parbox[t]{3em}{%
		\@tempcnta\@ne\relax
		\loop{\underline{\scriptsize\the\@tempcnta}}\\
		\advance\@tempcnta by \@ne\ifnum\@tempcnta<48\repeat}}

\begin{document}
	
	\title{Distributions-oriented wind forecast verification by a
		hidden Markov model for multivariate circular-linear data
	}
	
	
	\author[1]{Gianluca Mastrantonio} 
	\author[2]{Alessio Pollice}
	\author[3]{Francesca Fedele}
	\affil[1]{Dipartimento di Scienze Matematiche, Politecnico di Torino, Corso Duca degli Abruzzi, 24, 10129 Turin Italy}
	\affil[2]{ Dipartimento di Scienze Economiche e Metodi Matematici,
		Universit\`a degli Studi di Bari Aldo Moro, Largo Abbazia Santa Scolastica, 70124, Bari, Italy}
	\affil[3]{Agenzia Regionale per la Prevenzione e la Protezione dell'Ambiente (ARPA)
		Puglia, Corso Trieste 27, 70126, Bari, Italy}
	
	\date{}

	\maketitle

	\begin{abstract}
		Winds from the North-West quadrant and lack of precipitation are known to lead to an increase of PM10 concentrations over a residential
		neighborhood in the city of Taranto (Italy). In 2012 the local government prescribed a reduction of industrial emissions by 10\% every time
		such meteorological conditions are forecasted 72 hours in advance. Wind forecasting is addressed using the Weather Research and Forecasting
		(WRF) atmospheric simulation system by the Regional Environmental Protection Agency. In the context of distributions-oriented forecast
		verification, we propose a comprehensive model-based inferential approach to investigate the ability of the WRF system to forecast the local wind speed
		and direction allowing different performances for unknown   weather regimes. Ground-observed and WRF-forecasted wind speed and direction at a relevant
		location are jointly modeled as a 4-dimensional time series with an unknown finite number of states characterized by homogeneous distributional
		behavior. The proposed model relies on
		a mixture of joint projected and skew normal distributions with time-dependent states, where the temporal evolution of the state membership follows a first
		order Markov process. Parameter estimates, including the number of states, are obtained by a
		Bayesian MCMC-based method. Results provide useful insights on the performance of WRF forecasts in relation to different combinations of wind
		speed and direction.
	\end{abstract}

	\maketitle

\section{Introduction} \label{s:intro}
This work is concerned with a heavy industrial district located very
close to a residential area in the city of Taranto (Puglia region,
Italy) including the largest steel factory in Europe, an oil
refinery and a kiln cement factory. Emissions are mainly composed by suspended
particle matter, polycyclic aromatic hydrocarbon compounds and
benzene \citep{Fisherrr2003} and are associated to known adverse health effects \citep{Brunekreef2012}. In the last years, several PM10 limit
value exceedances (according to the 2008 European Air Quality Directive 2008/50/EC) were recorded
and these pollution events showed a close connection with
intense winds from North and North-West and lack of
precipitation, encouraging transportation from the industrial site
to the adjacent urban area \citep{Amodio2013,Fedele2014}. In 2012, the Puglia Local Government adopted a Regional Air Quality Plan
prescribing a reduction of industrial emissions by 10\% (with
respect to the daily mean values) every time such meteorological
conditions are forecasted 72 hours in advance. Here we focus on wind forecasting,
that the Puglia Environmental Protection Agency (ARPA
Puglia)
addresses by the Weather Research and Forecasting (WRF) mesoscale
numerical weather prediction system \citep{Skamarock2008,Tomasi2011,Fedele2015}. We are mainly concerned with
the investigation of the ability of the WRF system to properly
forecast winds blowing over the Gulf of Taranto. For this purpose we
consider hourly WRF-forecasted wind speed and direction data for year
2014 and the corresponding ground data collected at a specific
relevant point within the Tamburi neighborhood.


Verification is one aspect of measuring the quality of weather forecasts by comparison to relevant observations.
The traditional \textit{measures-oriented} approach to forecast verification involves obtaining a (generally small)
number of performance measures based on the posterior evaluation of a sample of past forecasts and observations. 
The alternative \textit{distributions-oriented}
approach acknowledges the intrinsic inferential nature of forecast verification and addresses the measure of
the quality of weather forecasts as an estimation problem, allowing for the explicit consideration of the
main sources of uncertainty involved in the process \citep{jolliffe2012forecast}.
The development of verification schemes based on the joint probability distribution of forecasts and observations has been urged by many authors
in the past \citep{Murphy87,Brooks1996}, as they allow proper investigation of the stochastic nature of the relationship between forecasts and observations
providing insights into strengths and weaknesses of the forecasting systems and showing areas where improvements can be obtained.
Parametric probability distributions are only occasionally assumed for this joint distribution in continuous
settings \citep[see][and references therein]{Wilks2011}. This is even more true in the investigation of the performance of wind field
forecasts, where observed and forecasted wind speed and direction form a continuous 4-dimensional mixed circular-linear variable.
Atmospheric simulation systems such as WRF show different performances for different weather  conditions
\citep{lefevre2010weather,rostkier2014objective,raktham2015simulation}. Then, for the purpose of investigating WRF forecast performance, we
consider the 4-dimensional time series of observed and forecasted wind speed and direction as characterized by an unknown number of
homogeneous states. Such states reproduce associated observed and forecasted wind conditions, accounting for the relation between wind
speed and direction.

There is a growing interest in circular data analysis, with examples arising in areas such as Oceanography
\citep{mastrantonio2015b,mastrantonio2015c,lagona2015}, Biology \citep{Maruotti2015,Hokimoto2014,Langrock2012} and Social Sciences.
\citep{Gill2010}. To our knowledge, the first joint circular-linear probability distribution model for more than two random variables was recently introduced by \cite{mastrantonio2015g}, namely the joint projected and skew normal (JPSN). Among the features that make
the JPSN attractive is the great flexibility and the possibility to introduce dependence between and within circular and linear variables.
In order to properly represent  homogeneous combinations of observed and forecasted wind conditions, we jointly model the time evolution of the
4-dimensional JPSN mixed variable by a hidden Markov model (HMM), 
i.e. a  mixture model with time-dependent states, where the state membership evolves according to a first order Markov process.
We adopt a non-parametric Bayesian estimation framework, allowing to estimate the unknown number of latent states considering Dirichlet process priors for transition
probabilities.
HMMs have already been used to analyze ground-observed wind speed and direction without forecast verification purposes. Most of the time independence between speed and
direction is assumed, as in \cite{Holzmann2006}, \cite{zucchini2009b}, \cite{Bulla2012}  and \cite{bulla2015}, but exceptions exist. For example
\cite{lagona2015} use the recently introduced Abe-Ley cylindrical density \citep{Abe2015} and \cite{mastrantonio2015} adopt the general circular-linear
projected normal.
In all the previous proposals the unknown number of latent states is not estimated, but it is rather assumed fixed a-priori.

In this proposal we define a comprehensive model-based approach that allows addressing distributions-oriented wind forecast
verification properly accounting for the circular nature of wind direction data. While the joint behavior of observed and forecasted wind
speed and direction is described by the JPSN, the HMM provides the representation of the time evolution with changing homogeneous
states. Coupling the JPSN with the HMM we obtain a flexible model and a rich parametrization that reproduce the relevant observed and forecasted
processes. Overall, computational efficiency characterizes the Bayesian MCMC estimation of the proposed model: HMM and JPSN parameters are both obtained by Gibbs sampling steps.

The remaining part of the paper is structured as follows. The environmental problem is stated in Section \ref{sec:data} where we report a
brief description of the relevant meteorology of the Taranto area and provide information on several data features. In Section
\ref{sec:projsekw} the JPSN probability distribution for multivariate mixed circular-linear random vectors is
introduced; some distributional features and implementation issues are also outlined. Section \ref{sec:HPD-HMM} is dedicated to the
definition of the HMM for circular-linear multivariate time series and to a brief discussion on the implementation of the relative Bayesian
inferences. Finally, in Section \ref{sec:res} we report the results for the Taranto case-study, highlighting the major advantages of the proposed methodology. Additional information and supporting material is available online at the journal's website.

\section{Observed and simulated wind data}\label{sec:data}

\begin{figure} [t]
    \centering
    \includegraphics[scale=0.4]{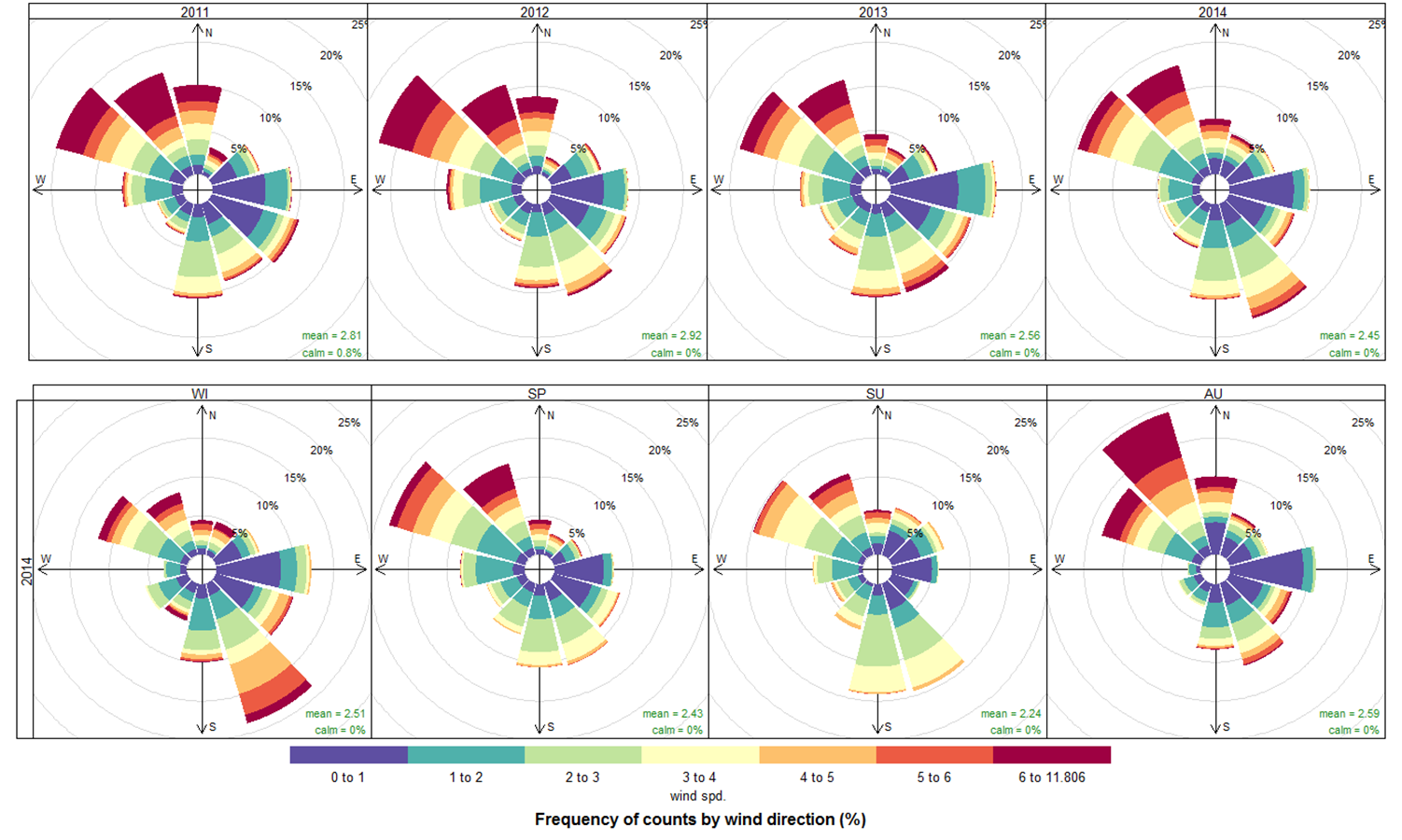}
    \caption{ Annual (top panel) and seasonal (bottom panel) wind roses at the San Vito ground station for 2011-2014 and 2014, respectively.} \label{fig:fig3}
\end{figure}

As already remarked in the introduction, we are concerned with the Tamburi neighborhood within the city of Taranto, here represented by the
position of the San Vito air quality monitoring station.
%
%
The San Vito ground station belongs to the ARPA Puglia air quality monitoring network, it is provided with six meteorological sensors
conform to the World Meteorological Organization standards and collects hourly wind speed and direction data since February 2002.
 The location of the San Vito monitoring station is characterized by an extreme proximity to the Ionian Sea. This makes wind measurements strongly affected
by land-sea breezes \citep{Stull1988} due to differences in the heat capacity and molecular conductivity between land surface and sea.
Though a direction is still associated to winds observed with low speed ($<0.3$ m/s), it should be noted that these measures are affected
by high variability. However, due to the proximity to the sea, the land-sea breeze effect causes the observed series to have no null wind
speed recordings and a very low percentage of small ones ($<4$\%). Missing values (2.1\% for wind direction and 2.0\% for wind speed) are
generally due to baffling winds or data transmission errors. 
%
A preliminary characterization of Taranto winds based on San Vito data is obtained by the annual wind roses for the period 2011-2014 and by
the seasonal wind roses for the year 2014 reported in Figure \ref{fig:fig3}.
In the annual roses (\ref{fig:fig3}, top) winds blowing from North-West are generally associated to high speed ($>$ 6m/s), while in the
other quadrants weaker winds are found. The average annual wind speed values range from 2.45 m/s in 2014 to 2.92 m/s in 2012. The seasonal
wind roses (\ref{fig:fig3}, bottom) show that North-West winds are prevailing in autumn and spring while in winter and summer winds from
the South-East quadrant are more frequent. Winds blowing from the North-West quadrant are stronger in autumn and spring, while the speed of
winds from South-East is higher in the colder than in the warmer seasons.
As a matter of fact, the intensity and evolution of winds in the Mediterranean area mainly depend on the position and extension of the
Azores and Siberian Anticyclones which in turn have more or less fixed seasonal configurations \citep{fantauzzo}. Therefore a highly
differential behavior of wind regimes is expected between the four seasons. Given the purpose of the present study, observed and forecasted
annual time series were partitioned into their seasonal components.

Hourly wind speed and direction forecasts are obtained 72 hours in advance for the whole year 2014 by the WRF atmospheric simulation
system. WRF is developed by a collaboration of research centers, universities and government agencies coordinated by the US National Center
for Atmospheric Research (NCAR).
Among the advantages of WRF is
the high flexibility that allows tuning physical parameterizations according to specific interests.
Predictive simulations are obtained by the ARW (Advanced Research WRF) core of the WRF system, solving the fully compressible
non-hydrostatic Euler equations using terrain-following hydrostatic-pressure vertical coordinates \citep{skamarock2005description} and the
Runge-Kutta integration method \citep{Butcher1987}. Specific WRF settings used in the implementation of this work include the following:
\begin{itemize}
    \item
    Among the different parametrization schemes offered by WRF to simulate physical processes, those used for radiation processes, surface
    processes, planetary boundary layer physics, cumulus processes and microphysics processes are listed in Table \ref{tab:tab1}.
    \item
    A one-way nesting configuration was chosen including two domains
    with different spatial resolution: a parent domain with 16 km grid
    spacing covering the central Mediterranean and a nested domain with
    4 km grid spacing covering the Puglia region.
    \item
    The default US Geological Survey (USGS) land cover database was replaced by the European CORINE land cover database
    \citep{Bossard2000,Buttner2004}, characterized by higher resolution and updated categories.
    \item
    The WRF software architecture allowed the use of parallel computing and part of the simulation process was run as a distributed memory job, with big advantages in reducing computation times. WRF was run on a high-performance computational infrastructure, made available within the ReCaS project, funded by the Italian Ministry of Education, University and Research.
\end{itemize}

\begin{table}
    \small
    \centering
    \begin{tabular}{l|l}
        Physics process & WRF scheme name\\ \hline \hline
        Radiation Processes&rrtm scheme \citep{Mlawer1997} and      \\ & Dudhia scheme \citep{Dudhia1989}\\
        Surface Processes&Noah Land Surface Model \citep{Chen2001}\\
        Planetary Boundary Layer Physics&Yonsei University scheme \citep{Hong2006}\\
        Cumulus Processes&Kain-Fritsch scheme \citep{Kain2004}\\
        Microphysics Processes&Thompson scheme \citep{Thompson2004,Thompson2008} \\ \hline
    \end{tabular}
    \caption{Summary of the WRF Physics parametrization schemes used in this study} \label{tab:tab1}
\end{table}

Simulations cover the period 3 January - 20 December 2014, with output temporal resolution fixed to one hour and 72 hours forecast time.
For comparability reasons, WRF-forecasted wind fields were obtained at the point location of the San Vito ground monitoring station.
 Missing values of the WRF-simulated series (4.5\%) are related to entire days for which the initial and boundary conditions are not
available. 

\section{The joint projected and skew normal distribution} \label{sec:projsekw}

As multivariate circular-linear distribution model for ground-observed and WRF-simulated wind speed and direction we consider the recently
introduced JPSN \citep{mastrantonio2015g}. In this Section we define the JPSN, discuss some identifiability issues,
show how to perform Bayesian inference in the i.i.d. case and provide statistics describing the main distributional features. The extension of the MCMC algorithm from the i.i.d. to the time series framework
is given in Section \ref{sec:HPD-HMM} together with the definition of the HMM used to represent the time evolution characterized by
homogeneous latent states.

Let $\mathbf{W}$ and $\mathbf{Y}$ be two real-valued random vectors of length $2p$ and $q$, respectively. As a first step
in the constructive definition of the JPSN we assume that the $(2p+q)$-dimensional random vector $(\mathbf{W},\mathbf{Y})^{\prime}$ is
distributed as a multivariate \emph{skew normal} distribution \citep[hereafter SN,][]{sahu2003} with parameters $\boldsymbol{\mu}$,
$\boldsymbol{\Sigma}$ and $\text{diag}{((\mathbf{0}_{2p},\boldsymbol{\lambda})')}$: $(\mathbf{W},\mathbf{Y})^{\prime}\sim
\text{SN}_{2p+q}(\boldsymbol{\mu}, $ $ \boldsymbol{\Sigma},$ $\text{diag}((\mathbf{0}_{2p},\boldsymbol{\lambda})'))$, where
$\boldsymbol{\mu} = (\boldsymbol{\mu}_w,\boldsymbol{\mu}_y)' \in\mathbb{R}^{2p+q}$, $\boldsymbol{\Sigma}=\left(
\begin{array}{cc}
\boldsymbol{\Sigma}_{w} & \boldsymbol{\Sigma}_{wy}\\
\boldsymbol{\Sigma}_{wy}^{\prime} & \boldsymbol{\Sigma}_{y}
\end{array}
\right)$ is a non-negative definite $(2p+q)\times (2p+q)$ matrix, $\mathbf{0}_{2p}$ is a vector of zeros of length $2p$,
$\boldsymbol{\lambda}=\{ \lambda_{j} \}_{j=1}^q\in\mathbb{R}^{q}$ and $\mbox{diag}(\cdot)$ indicates a diagonal matrix with
non-null elements given by its argument. The SN distribution introduces skewness in the multivariate normal through the so called
\emph{skew matrix}, in our case given by $\text{diag}((\mathbf{0}_{2p},\boldsymbol{\lambda})')$. 
Since we assume a diagonal skew matrix, the SN distribution is closed under marginalization \citep{sahu2003}, then $\mathbf{W} \sim
\text{N}_{2p}(\boldsymbol{\mu}_w, \boldsymbol{\Sigma}_w)$ and $\mathbf{Y}\sim \text{SN}_{q}(\boldsymbol{\mu}_y, \boldsymbol{\Sigma}_y,
\text{diag}(\boldsymbol{\lambda}))$. The SN distribution has the following alternative stochastic representation  \citep{li2005}, that
will
prove to be useful later in this section. 
In our notation, $(\mathbf{W},\mathbf{Y})^{\prime}\sim \text{SN}_{2p+q}(\boldsymbol{\mu}, \boldsymbol{\Sigma},
\text{diag}((\mathbf{0}_{2p},\boldsymbol{\lambda})'))$ implies:
\begin{align}
\mathbf{W} &= \boldsymbol{\mu}_w+\mathbf{H}_w \label{eq:skew1},\\
\mathbf{Y} & = \boldsymbol{\mu}_y+\text{diag}(\mathbf{D})  \boldsymbol{\lambda}+\mathbf{H}_y\label{eq:skew2},
\end{align}
where $(\mathbf{H}_w,\mathbf{H}_y)' \sim \text{N}_{2p+q}(\mathbf{0}_{2p+q},\boldsymbol{\Sigma})$ and $\mathbf{D} \sim
\text{HN}_q(\mathbf{0}_q,\mathbf{I}_q)$, where  $\mathbf{I}_q$ is the $q-$dimensional identity matrix and $\text{HN}_q$ indicates the
\emph{half normal} distribution, i.e. a truncated normal defined over $\{\mathbb{R}^q\}^+$.

As a second step in the definition of the JPSN distribution, we build a vector of  circular variables partitioning
$\mathbf{W}$ into $p$ couples $\mathbf{W}_i=(W_{i1},W_{i2})^{\prime}$ and transforming each  $\mathbf{W}_i$ using the following relation:
\begin{equation} \label{eq:tranpn}
    \Theta_i = \text{atan}^* \frac{W_{i2}}{W_{i1}} \in [0 ,2 \pi),
\end{equation}
where $i=1,\ldots,p$ and $\text{atan}^*$ is a modified  arctangent function  \citep{Jammalamadaka2001}. Equation
\eqref{eq:tranpn} transforms the normally distributed 2-dimensional random variable $\mathbf{W}_i$ into a circular variable $\Theta_i$ with
\emph{projected normal} distribution \citep{Wang2013}. Notice that also the following relations hold:
    \begin{align}
    W_{i1} &=  R_i\cos \Theta_i, \label{eq:theta1}\\
    W_{i2} &=  R_i\sin \Theta_i,\label{eq:theta2}
    \end{align}
where $R_i=||\mathbf{W}_i||$.

Following \cite{mastrantonio2015g}, the vector of $p$ circular and $q$ linear variables
$(\boldsymbol{\Theta}, \mathbf{Y})^{\prime}$, obtained transforming $(\mathbf{W}, \mathbf{Y})^{\prime}$ by \eqref{eq:tranpn} is said to
have
$(p,q)$-variate \emph{joint projected and skew normal} distribution with parameters $\boldsymbol{\mu}$, $\boldsymbol{\Sigma}$ and
$\boldsymbol{\lambda}$: $(\boldsymbol{\Theta},\mathbf{Y})^{\prime}\sim \text{JPSN}_{p,q}(\boldsymbol{\mu}, \boldsymbol{\Sigma},
\boldsymbol{\lambda})$.
The JPSN distribution does not have a closed form expression, but the joint density of the ``augmented'' vector
$(\boldsymbol{\Theta},\mathbf{Y},{\mathbf{R}},{\mathbf{D}})'$, with $\mathbf{R} = \{R_i\}_{i=1}^p$, is easily obtained using \eqref{eq:skew1},
\eqref{eq:skew2}, \eqref{eq:theta1} and \eqref{eq:theta2}:
\begin{align}
&   f(\boldsymbol{\theta},\mathbf{y},{\mathbf{r}},{\mathbf{d}}|{\boldsymbol{\mu}}, {\boldsymbol{\Sigma}},\boldsymbol{\lambda})= \\
&   2^q \phi_{2p+q}((\mathbf{w}, \mathbf{y} )^{\prime}| (\boldsymbol{\mu}_w,\boldsymbol{\mu}_y+\text{diag}(\mathbf{d})
 \boldsymbol{\lambda}), \boldsymbol{\Sigma})  \phi_q(\mathbf{d}| \mathbf{0},\mathbf{I}_q ) \prod_{i=1}^p r_{i},\label{eq:apsn}
\end{align}
where  $\phi$ is the multivariate normal probability density function. Vector $(\boldsymbol{\Theta}, \mathbf{Y},
\mathbf{R},\mathbf{D})^{\prime}$ is said to be distributed as an \emph{augmented joint projected and skew normal}: $(\boldsymbol{\Theta},
\mathbf{Y},\mathbf{R}, \mathbf{D})^{\prime}\sim $ $ \text{AugJPSN}_{p,q}(\boldsymbol{\mu},$ $\boldsymbol{\Sigma},\boldsymbol{{\lambda}})$.
As marginalization of \eqref{eq:apsn} over  $\mathbf{R}$ and $\mathbf{D}$ gives the JPSN density, the JPSN parameter
estimation algorithm is based on the AugJPSN, treating the elements of $({\mathbf{R}},{\mathbf{D}})'$ as latent variables.

\subsection{Identifiability of the JPSN}
The projected normal, in its general form, is known to have an identifiable issue  \citep{Wang2013}. Let
$\mathbf{C}_w= \mbox{diag}(\{(c_i,c_i)\}_{i=1}^p)$, with $c_i \in \mathbb{R}^+$, notice that the two random vectors $\mathbf{W} \sim
N_{2p}(\boldsymbol{\mu}_w,\boldsymbol{\Sigma}_w)$ and $ \mathbf{C}_w \mathbf{W} \sim N_{2p}(\mathbf{C}_w
\boldsymbol{\mu}_w,\mathbf{C}_w\boldsymbol{\Sigma}_w\mathbf{C}_w)$ 
produce the same $\boldsymbol{\Theta}$  since $c_i$'s cancel out in equation \eqref{eq:tranpn}. Then
$PN_p(\boldsymbol{\mu}_w,\boldsymbol{\Sigma}_w)$  and $PN_p(\mathbf{C}_w\boldsymbol{\mu}_w,\mathbf{C}_w\boldsymbol{\Sigma}_w\mathbf{C}_w)$
are the same distribution and the model is not identifiable. The JPSN suffers from the same problem since the marginal distribution of
its circular component is the projected normal. Let $\mathbf{C}= \mbox{diag}\left( \left\{c_i,c_i  \right\}_{i=1}^p ,\mathbf{1}_q\right)$, then 
$\mbox{JPSN}_{p,q}(\boldsymbol{\mu},\boldsymbol{\Sigma}, \boldsymbol{\lambda})$ and
$\mbox{JPSN}_{p,q}(\mathbf{C}\boldsymbol{\mu},\mathbf{C}\boldsymbol{\Sigma}\mathbf{C}, \boldsymbol{\lambda})$ are the same distribution.
As a consequence, the model is not identifiable unless some constraints  are adopted.

Setting the variance of $W_{i2}$ to 1 ($i=1,2,\dots,p$)  addresses the identification problem \citep{Wang2013}, but the constraints hamper the estimation of $\boldsymbol{\Sigma}$, due to the unavailability of algorithms for constrained covariance matrix estimation. Alternatively, let
$\sigma_{w_{i2}}^2$ be the variance of $W_{i2}$ and $c_i=1/\sigma_{w_{i2}}$ and notice that
$(\mathbf{C}\boldsymbol{\mu},\mathbf{C}\boldsymbol{\Sigma}\mathbf{C})$ and $(\boldsymbol{\mu}, \boldsymbol{\Sigma})$ produce the same JPSN
density but, by construction, $\mathbf{C}\boldsymbol{\Sigma}\mathbf{C}$ complies with the identifiability constraints. The algorithm proposed by \cite{mastrantonio2015g} obtains posterior samples from the  non-identifiable model
and re-scales each sample of  $(\boldsymbol{\mu},\boldsymbol{\Sigma})$ to the identifiable version
$(\mathbf{C}\boldsymbol{\mu},\mathbf{C}\boldsymbol{\Sigma}\mathbf{C})$.

\subsection{JPSN estimation: MCMC implementation details}

Posterior samples of JPSN parameters are easily obtained by the augmented representation of the JPSN in \eqref{eq:apsn},
under suitable prior choices. More precisely, for a set of $T$ i.i.d. observations $(\boldsymbol{\Theta}_t,\mathbf{Y}_t )^{\prime}\sim
\text{JPSN}_{p,q}(\boldsymbol{\mu},$ $\boldsymbol{\Sigma},\boldsymbol{{\lambda}})$, $t=1,\dots,T$
the joint full conditional of JPSN parameters $\boldsymbol{\mu}$, $\boldsymbol{\Sigma}$ and $\boldsymbol{\lambda}$ is proportional to
\begin{equation} \label{eq:cond}
\prod_{t=1}^T \phi_{2p+q}((\mathbf{w}_t, \mathbf{y}_t )^{\prime}| (\boldsymbol{\mu}_w,\boldsymbol{\mu}_y+\text{diag}(
\mathbf{d}_t)\boldsymbol{\lambda})', \boldsymbol{\Sigma}) f(\boldsymbol{\mu},\boldsymbol{\Sigma},\boldsymbol{\lambda}).
\end{equation}
This latter expression is also obtained as the joint full conditional for a multivariate normal likelihood, where the mean depends on $\boldsymbol{\mu}$, $\mathbf{d}_t$ and $\boldsymbol{\lambda}$, with a given prior $f(\cdot)$ over $\boldsymbol{\mu}$, $\boldsymbol{\Sigma}$ and $\boldsymbol{\lambda}$. Notice that $\boldsymbol{\mu}$, $\mbox{diag}(\mathbf{d}_t)$ and $\boldsymbol{\lambda}$ respectively play the roles of intercept,  design matrix and regression coefficients in a Bayesian regression framework.
 Then standard priors used in this context can be used to
implement Gibbs-based MCMC steps. As suggested by \cite{mastrantonio2015g}, assuming a normal inverse-Wishart (NIW) prior
for $(\boldsymbol{\mu},\boldsymbol{\Sigma})$ and an independent normal prior for $\boldsymbol{\lambda}$, the MCMC algorithm can
conveniently be based on Gibbs steps. In fact in this case the full conditionals of $(\boldsymbol{\mu},\boldsymbol{\Sigma})$ and
$\boldsymbol{\lambda}$ are still respectively NIW and normal, the one of $\mathbf{d}_t$ is truncated normal and $\mathbf{r}_t$
can be simulated using the slice-sampling strategy proposed by \cite{Stumpfhause2016}.

\subsection{Statistics for JPSN distributional features}

JPSN parameters $\boldsymbol{\mu}_{y}$,  $\boldsymbol{\Sigma}_y$ and $\boldsymbol{\lambda}$ have a straightforward interpretation, since
\eqref{eq:skew1} and \eqref{eq:skew2} imply that
\begin{equation}
E(\mathbf{Y}) =\boldsymbol{\mu}_{y}+ \sqrt{\frac{2}{\pi}} \boldsymbol{\lambda},\qquad \mbox{Var}(\mathbf{Y}) =\boldsymbol{\Sigma}_{y}+
\left(1- \frac{2}{\pi}\right)  \mbox{diag}(\boldsymbol{\lambda} \boldsymbol{\lambda}^{\prime}), \label{eq:lmeanvar}
\end{equation}
and $\boldsymbol{\lambda}$ controls the skewness of the distribution of the linear component $\mathbf{Y}$. Matrix-valued parameters
$\boldsymbol{\Sigma}_{wy}$ and $\boldsymbol{\Sigma}_{w}$ control the circular-linear and circular-circular dependence since $\mathbf{W}_i
\perp  \mathbf{W}_j$ implies $\Theta_i \perp \Theta_j$ and $\mathbf{W}_i \perp  Y_j$ implies $\Theta_i \perp Y_j$, where $\perp$ indicates
independence. It is not very clear how all parameters jointly influence the density of $(\boldsymbol{\Theta},\mathbf{Y})^{\prime}$, however
Monte Carlo (MC) approximations of the main features of the JPSN distribution are obtained in the Bayesian estimation framework, bypassing
the afore mentioned difficulties. MC approximations of the circular mean and concentration of $\Theta_{i}$ are respectively obtained
sampling the following functions of $\Theta_{i}$:
\begin{equation}
\alpha_{i} = \mbox{atan}^* \frac{E(\sin \Theta_{i} )}{E(\cos \Theta_{i} )}\label{eq:cmean}
\end{equation}
and
\begin{equation}
\zeta_{i} = \sqrt{E(\sin \Theta_{i} )^2+E(\cos \Theta_{i} )^2}.\label{eq:cconc}
\end{equation}
A measure of the correlation between circular variables \citep{fisher1996} taking values in $[-1,1]$ is given by
\begin{equation}
\rho_{(\Theta_i,\Theta_{i^{\prime}})} =    \frac{E(\sin(\Theta_i-\Theta_i^*)\sin(\Theta_{i^{\prime}}-\Theta_{i^{\prime}}^*)    )}{\sqrt{
E(\sin^2(\Theta_i-\Theta_i^*)) E(\sin^2(\Theta_{i^{\prime}}-\Theta_{i^{\prime}}^*)   )     }} ,\label{eq:112}
\end{equation}
where the bivariate random variable $(\Theta_i^*,\Theta_{i^{\prime}}^*)$ is distributed as $(\Theta_i,\Theta_{i^{\prime}})$. A
circular-linear dependence measure taking values in $[0,1]$
\citep{mardia1976} is given by:
\begin{align}
\rho_{(\Theta_i,Y_{j})}^2 = & \frac{\text{Cor}{(\cos\Theta_i,Y_{j} )}^2+\text{Cor}{(\sin\Theta_i,Y_{j} )}^2}{1-\text{Cor}{(\cos\Theta_i,\sin\Theta_i )}}-\\
& \frac{2\text{Cor}{(\cos\Theta_i,Y_{j}
        )}\text{Cor}{(\sin\Theta_i,Y_{j} )}\text{Cor}{(\cos\Theta_i,\sin\Theta_i )}}{1-\text{Cor}{(\cos\Theta_i,\sin\Theta_i )}} . \label{eq:111}
\end{align}

\section{A hidden Markov model for observed and forecasted wind fields} \label{sec:HPD-HMM}

In this section we introduce the joint model for observed and forecasted wind speed and direction that was estimated for
each of the four seasons of year 2014. We indicate the ground-observed and WRF-simulated wind direction and log speed at time $t$ with
$(\Theta_{tg},\mathbf{Y}_{tg} )'$ and $(\Theta_{ts},\mathbf{Y}_{ts} )'$, respectively, assuming $t=1,2,\dots , T$. Notice that in this case
$p=q=2$.

 To catch the most relevant interactions between observed and forecasted wind speed and direction, the 4-dimensional circular-linear
time series is assumed to be characterized by homogeneous states and is modeled by a mixture of JPSN
distributions with time-dependent states, where the temporal evolution of the state membership follows a first order Markov process, namely
a HMM. The HMM is estimated within a non-parametric Bayesian framework, relying on Dirichlet process priors for transition probabilities,
thus leading to a multivariate circular-linear version of the sticky hierarchical Dirichlet process HMM (sHDP-HMM) proposed by
\cite{fox2011}. This specification allows us to estimate the unknown number of latent states, along with all other model parameters.

As was mentioned in Section \ref{sec:projsekw}, the latent variables $(R_{tg},R_{ts},D_{tg},D_{ts})'$ need to be
introduced to estimate JPSN parameters, leading to the augmented JPSN. Overall, manifest and latent observables are:
$\boldsymbol{\Theta}_t=(\Theta_{tg},\Theta_{ts})'$, $\mathbf{Y}_t=(\mathbf{Y}_{tg},\mathbf{Y}_{ts})'$, $\mathbf{R}_t=(R_{tg},R_{ts})'$ and
$\mathbf{D}_t=(D_{tg},D_{ts})'$.
At times $t=1,\ldots,T$ let $z_t \in  \mathbb{N}$ be a discrete random variable that represents the state of the HMM. Let
$\boldsymbol{\psi}_k = (\boldsymbol{\mu}_k,\boldsymbol{\Sigma}_k, \boldsymbol{\lambda}_k)^{\prime} $ be the set of  JPSN parameters for
state $k$ and $\boldsymbol{\pi}_{k}=\{\pi_{kj} \}_{j \in \mathbb{N}}$ be  the $k$-th row of the transition matrix, i.e. a probability vector. Then, letting
$\boldsymbol{\Theta}=\{\boldsymbol{\Theta}_t\}_{t=1}^T$, $\mathbf{Y} = \{ \mathbf{Y}_t \}_{t=1}^T$, $\mathbf{R} = \{ \mathbf{R}_t
\}_{t=1}^T$ and $\mathbf{D} = \{ \mathbf{D}_t \}_{t=1}^T$,  the sHDP-HMM
is given by:
\begin{align}
f(\boldsymbol{\theta},\boldsymbol{y},\mathbf{r},\mathbf{d}|\{z_t \}_{t=1}^T ,\{\boldsymbol{\psi}_{k}\}_{k \in \mathbb{N}}) & = \prod_{t=1}^T \prod_{k \in \mathbb{N}} f(\boldsymbol{\theta}_t,\mathbf{y}_t,\mathbf{r}_t,\mathbf{d}_t|\boldsymbol{\psi}_{k})^{I(z_t,k)}, \label{eq:hmm1}\\
\boldsymbol{\Theta}_t,\mathbf{Y}_t,\mathbf{R}_t, \mathbf{D}_t|\boldsymbol{\psi}_{k} &\sim \text{AugJPSN}_{2,2}(\boldsymbol{ \mu}_k, \boldsymbol{\Sigma}_k,\boldsymbol{\lambda}_k),\label{eq:hmm2}\\
z_{t}|z_{t-1},\{\boldsymbol{\pi}_k\}_{k \in \mathbb{N}} & \sim  \boldsymbol{\pi}_{z_{t-1}},\label{eq:hmm3}\\
\boldsymbol{\pi}_{{k}}  | \varsigma,\gamma,\{\beta_j \}_{j\in  \mathbb{N}}& \sim \text{DP}\left(\gamma,     (1-\varsigma) \{\beta_j \}_{j\in  \mathbb{N}}+\varsigma I(k,j)    \right) , \label{eq:distpi}\\
\{\beta_j\}_{j\in  \mathbb{N}}|\tau & \sim \text{GEM}(\tau) ,\label{eq:hdp3} \\
\boldsymbol{\psi}_{k}| H &\sim H, \, k \in  \mathbb{N} \label{eq:H}
\end{align}
where $\text{DP}(\cdot,\cdot)$ is the Dirichlet process, $\text{GEM}(\cdot)$ is the GEM distribution \citep{Pitman2006}, $\varsigma \in
[0,1]$, $\gamma>0$ and $\tau>0$ are hyperparameters, ${H}$ is a valid prior distribution over the domain of $\boldsymbol{\psi}_k$,
 $I(\cdot,\cdot)$ is the indicator function and $z_0=1$ is assumed \citep{cappe2005}. Clearly, marginalization over
$(\mathbf{R},\mathbf{D} )^{\prime}$   gives the sHDP-HMM for circular-linear data  $\{\boldsymbol{\theta}_t,\mathbf{y}_t\}_{t=1}^T$ with
$\boldsymbol{\Theta}_t,\mathbf{Y}_t|\boldsymbol{\psi}_{k} \sim \text{JPSN}_{2,2}(\boldsymbol{ \mu}_k,
\boldsymbol{\Sigma}_k,\boldsymbol{\lambda}_k)$  for the $k$-th mixture component.

Equations \eqref{eq:hmm1}, \eqref{eq:hmm2} and \eqref{eq:hmm3} define the standard HMM \citep{zucchini2009b}, where $\{z_t\}_{t=1}^T$ is
the latent discrete Markov process with the $\text{AugJPSN}$ as emission distribution. Notice that although the number of
components is potentially infinite in the specification of the sHDP-HMM, it is actually bounded by the length $T$ of the the observational
period. 
%
To give an intuitive interpretation of equations \eqref{eq:distpi}, let the latent state $k$ be \emph{non-empty} if at least one $z_t$ is
equal to $k$. Without loss of generality, let $\boldsymbol{\pi}_k^*=\left( \pi_{k1},\dots ,\pi_{kK},\sum_{j=K+1}^{\infty}\pi_{kj} \right)$, with $k\leq K$,
where the first $K$ states are non-empty. Then, following the standard definition of the DP \citep[see][]{sethuraman:stick}, equations
\eqref{eq:distpi} can equivalently be written as
\begin{align}
&\boldsymbol{\pi}_k^* |\varsigma,\gamma,\{\beta_j \}_{j\in \mathbb{N}},\boldsymbol{\psi}_{k}\sim \\
&\text{Dir}\left(\gamma((1-\varsigma)\beta_{1}+\varsigma I(k,1))      ,\dots,\gamma((1-\varsigma)\beta_{K}+\varsigma I(k,K) )    ,
\gamma(1-\varsigma)\sum_{j=K+1}^{\infty}\beta_j  \right) \label{eq:sw},
\end{align}
where $\text{Dir}(\cdot)$ is the Dirichlet distribution.  Then, as $E(\pi_{kj}) = (1-\varsigma)\beta_j+\varsigma I(k,j)$ and $Var(\pi_{kj})
= \frac{ ((1-\varsigma)\beta_{j}+\varsigma I(k,j)) (1- (1-\varsigma)\beta_{j}-\varsigma I(k,j))  }{\gamma+1}$, it follows that $\{\beta_j
\}_{j \in \mathbb{N}}$ and $\varsigma$ rule the mean and variance of $\boldsymbol{\pi}_k^*$, with  $\varsigma $ being an additional weight
added to the self-transition probability $\pi_{kk}$ to avoid the tendency to create redundant mixture components
\citep[see][]{teh2010,fox2011}.
The distribution of the $\beta_j$s is ruled by $\tau$ and concentrates its probability mass on fewer $\beta_j$s as $\tau$
decreases. Variable $K$ implicitly depends on parameters $\tau$, $\gamma$ and $\varsigma$ that 
have the following standard weak informative priors: $\tau \sim G(1, 0.01)$, $\gamma \sim G(1, 0.01)$ and
$\varsigma \sim B(1, 1)$, where $G(\cdot,\cdot)$ indicates the gamma distribution in terms of shape and scale and $B(\cdot, \cdot)$ is the
beta distribution. These priors allow to update the latent discrete time series $\{z_t\}_{t=1}^T$ and all parameters of the sHDP-HMM using
only Gibbs steps \citep[see][]{fox2011,Beal2002}.

To appreciate how the estimation algorithm proposed in Section \ref{sec:projsekw} for the i.i.d. case can be used in this
context, notice that observations in state $k$ are i.i.d. conditionally on the process $\{z_t \}_{t=1}^T $.
Then priors suggested for the i.i.d. case in Section \ref{sec:projsekw} allow to update parameters and latent variables using only Gibbs
steps. We use the following standard weak informative prior settings: $\boldsymbol{\mu}_k,\boldsymbol{\Sigma}_k \sim \text{NIW} \left(
\mathbf{0}_{6}, 0.001, 15, \mathbf{I}_{6}  \right)$, $\boldsymbol{\lambda}_k \sim N_{2}\left( \mathbf{0}_2, 100 \mathbf{I}_2 \right)$.

Notice that, despite the high complexity of the model, the MCMC algorithm needed to estimate the model unknowns is only based on Gibbs
steps.

\section{Results}\label{sec:res}

 The estimation algorithms specified in Sections \ref{sec:projsekw} and \ref{sec:HPD-HMM} were applied to the
four seasonal time series of hourly observed and forecasted wind speed and direction covering 77, 92, 92, 91 days, respectively. 
Prior to modeling, in order to comply with the domain of the JPSN, circular variables were transformed from degrees to radians and the log
of speed was taken for both ground-observed and WRF-simulated wind data. { Missing values were predicted along with all unknown parameters during model fitting.} To improve the interpretation of graphical and tabular displays 
the results were back-transformed to their original units. All models were estimated considering 400,000 iterations, with 300,000 for the
burn-in phase and thinning by 20, i.e. taking 5,000 samples for inferential purposes. For each season, our R/C++ implementation of the
estimation procedure took about 3 hours with with a 2.5 GHz Intel Core i5 processor. The number of states of the sHDP-HMM was estimated as
5 for all of the four seasons, with $P(K=5| \boldsymbol{\theta},\mathbf{y})\approx 1$. The five components were ordered, based on
increasing ground speed sHDP-HMM posterior means.
We want to stress that the latent sHDP-HMM states cannot be interpreted as wind regimes, since they jointly represent correlated
ground-observed and WRF-simulated winds that do not necessarily agree in terms of speed and direction. An example and some details are
given in the supplementary material.

\begin{figure}[t!]
    \captionsetup[subfigure]{labelformat=empty}
    \centering
    \subfloat[]{\raisebox{+0.75in}{\rotatebox[origin=f]{90}{WINTER}}}
    {\subfloat{\includegraphics[trim= {0.7cm 1.cm 0.7cm 1.2cm},clip,scale=0.30]{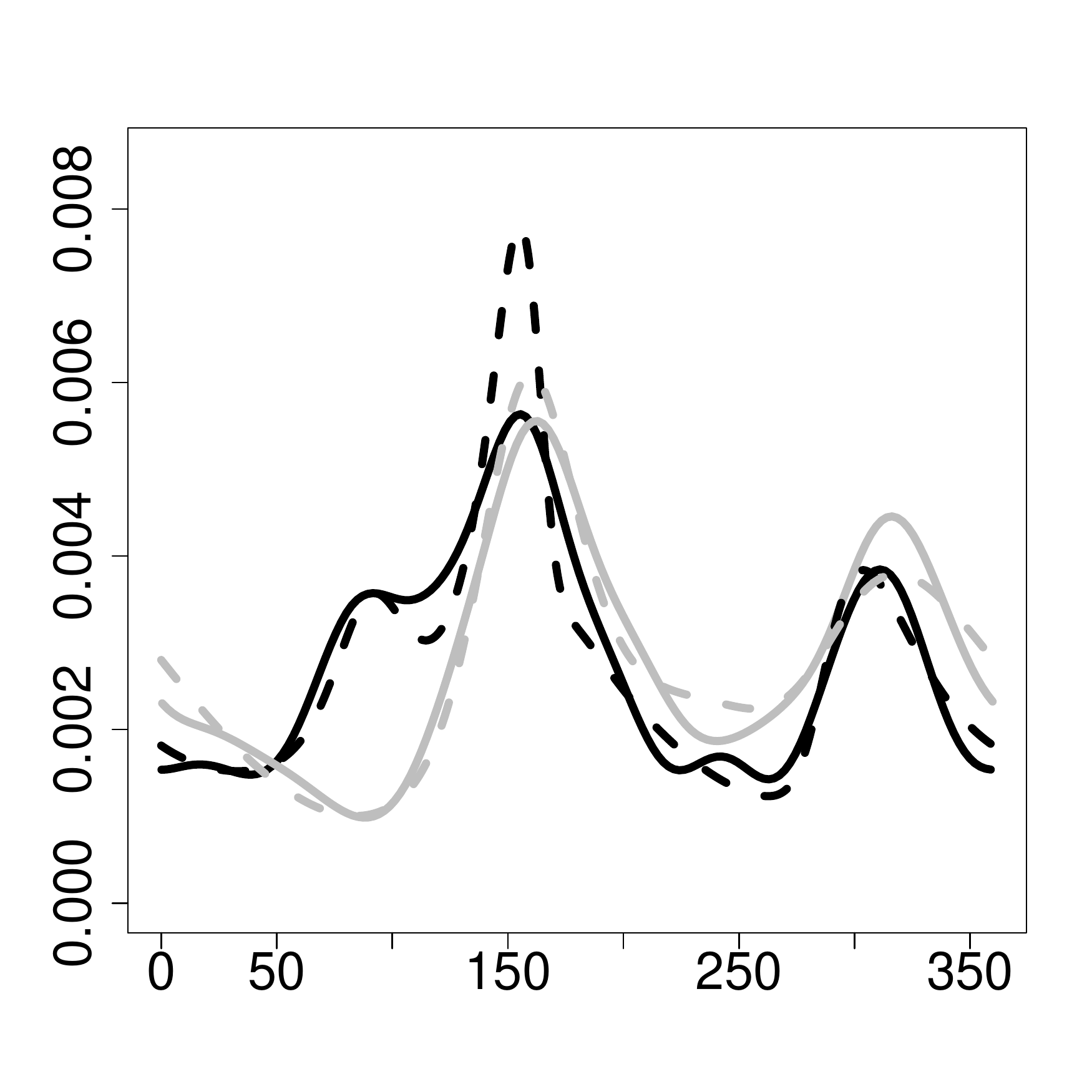}}}
        {\subfloat[]{\includegraphics[trim= {0.7cm 1.cm 0.7cm 1.2cm},clip,scale=0.30]{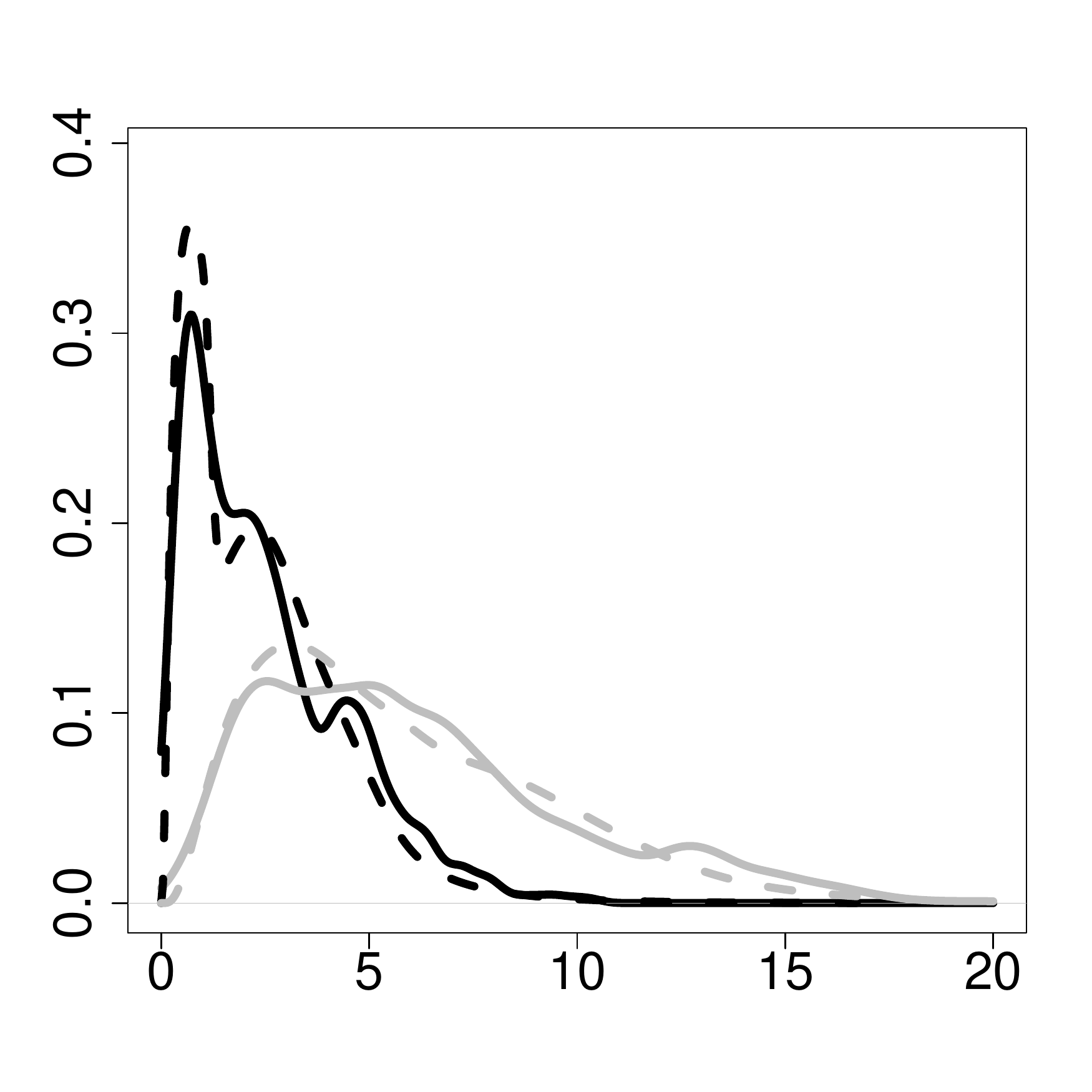}}}\\
    \subfloat[]{\raisebox{+0.75in}{\rotatebox[origin=f]{90}{SPRING}}}
    {\subfloat{\includegraphics[trim= {0.7cm 1.cm 0.7cm 1.2cm},clip,scale=0.30]{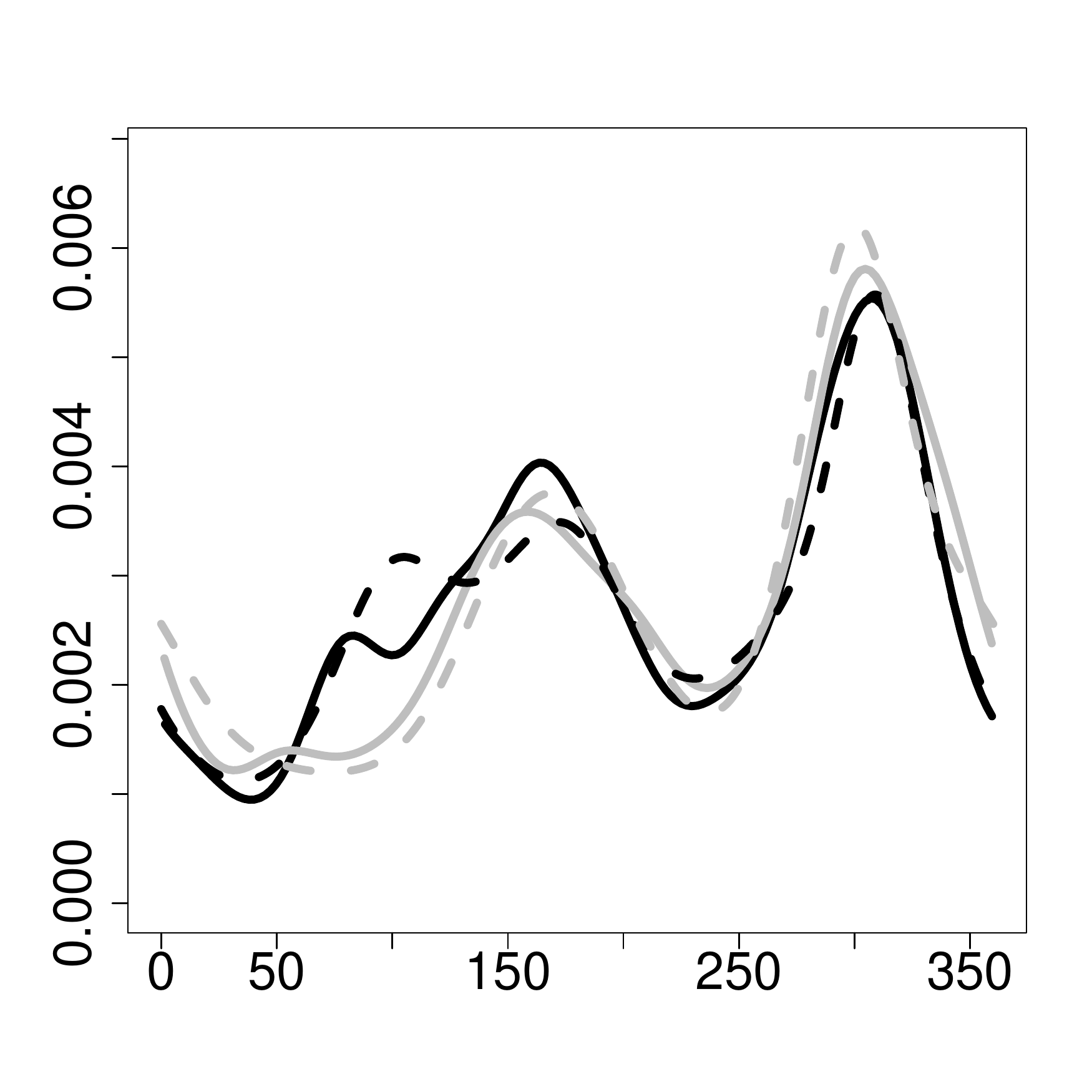}}}
    {\subfloat[]{\includegraphics[trim= {0.7cm 1.cm 0.7cm 1.2cm},clip,scale=0.30]{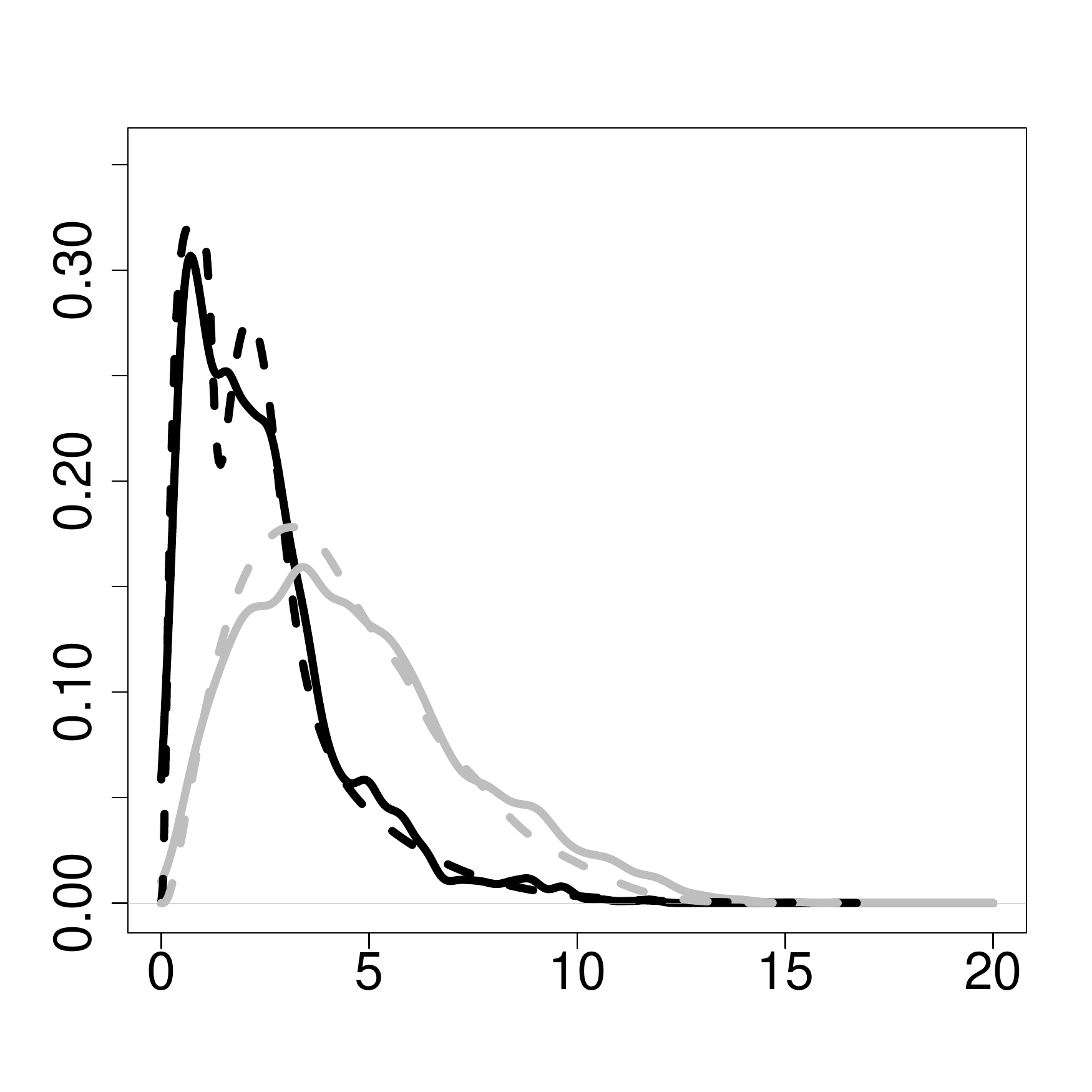}}}\\
    \subfloat[]{\raisebox{+0.75in}{\rotatebox[origin=f]{90}{SUMMER}}}
    {\subfloat{\includegraphics[trim= {0.7cm 1.cm 0.7cm 1.2cm},clip,scale=0.30]{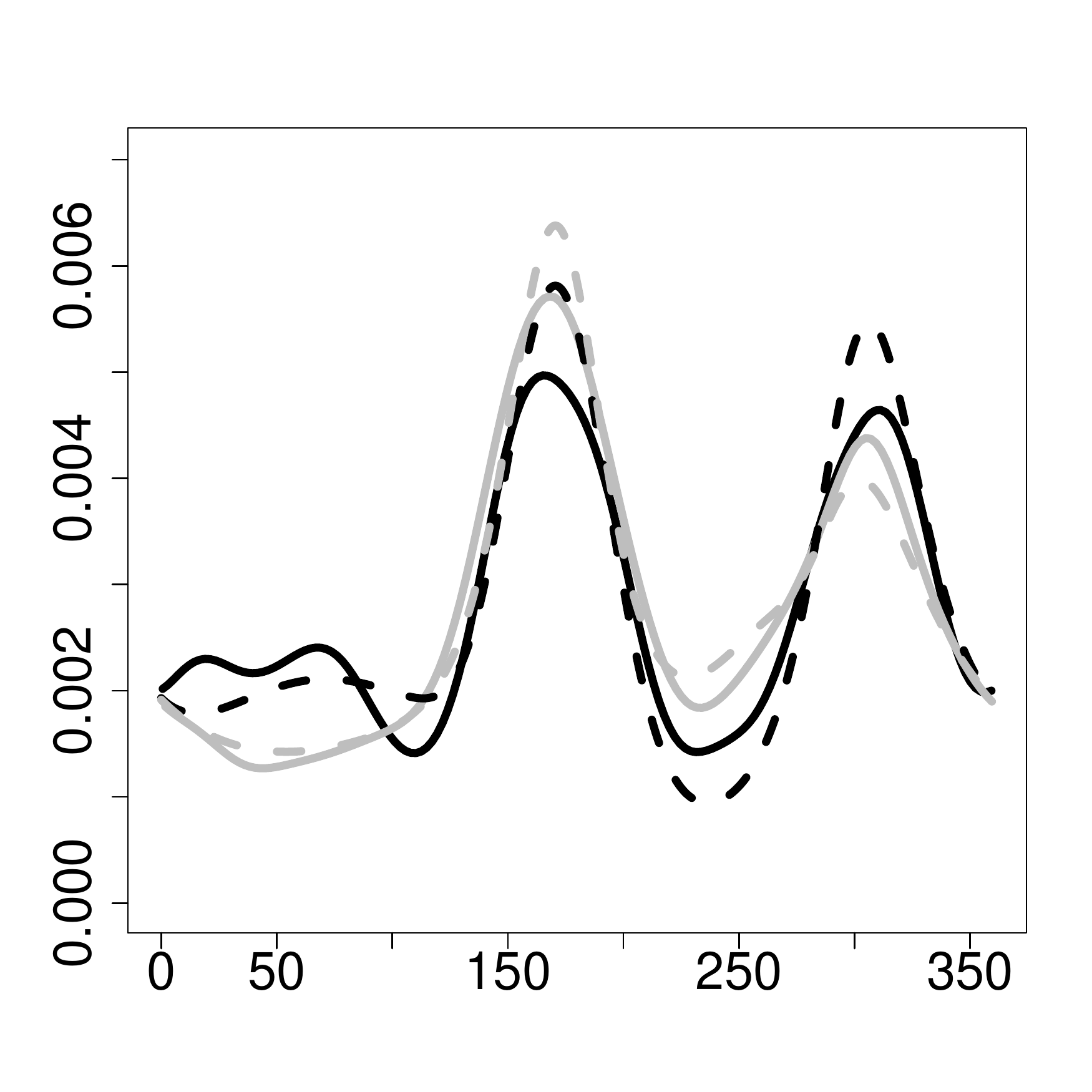}}}
    {\subfloat[]{\includegraphics[trim= {0.7cm 1.cm 0.7cm 1.2cm},clip,scale=0.30]{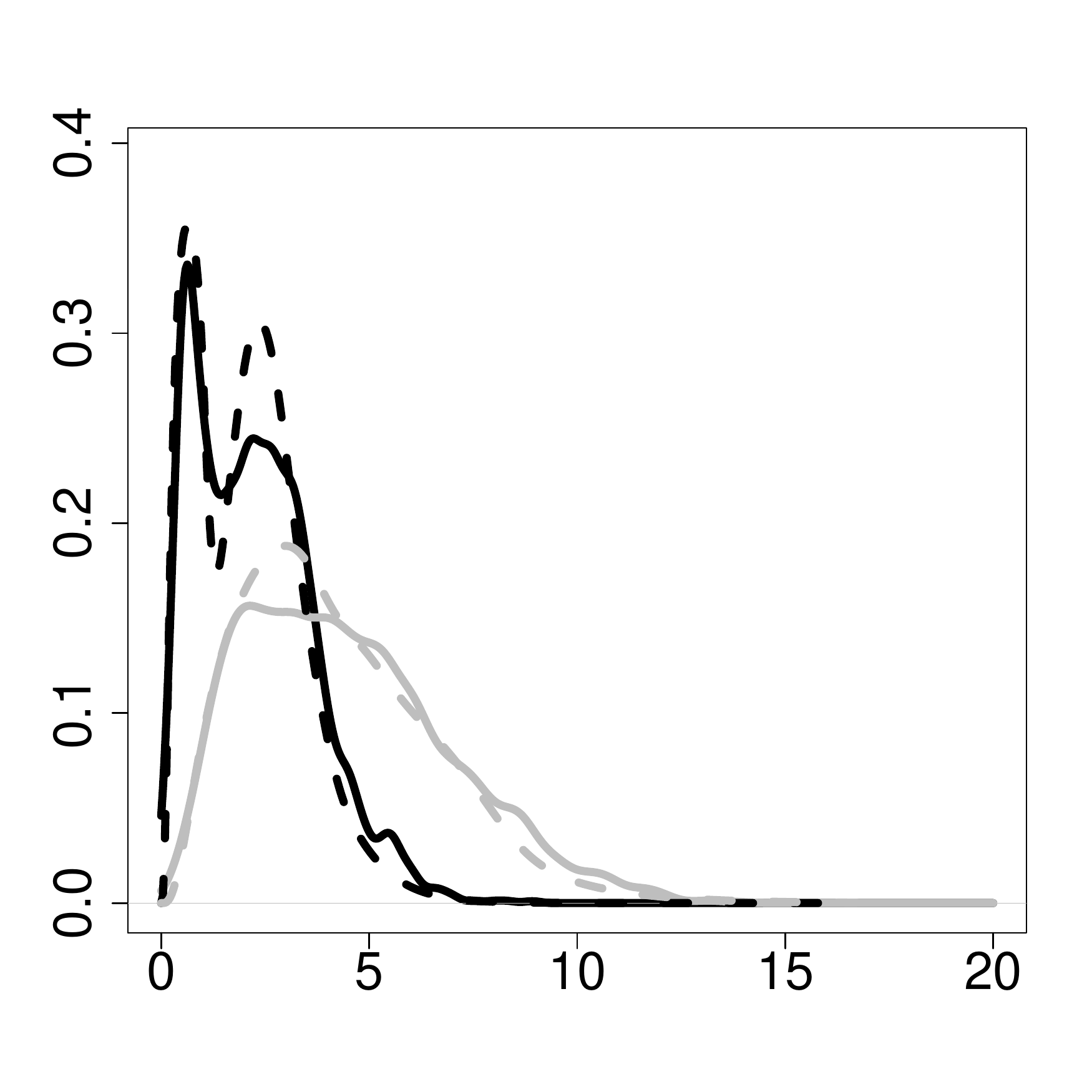}}}\\
    \subfloat[]{\raisebox{+0.75in}{\rotatebox[origin=f]{90}{AUTUMN}}}
    {\subfloat[WIND DIRECTION]{\includegraphics[trim= {0.7cm 1.cm 0.7cm 1.2cm},clip,scale=0.30]{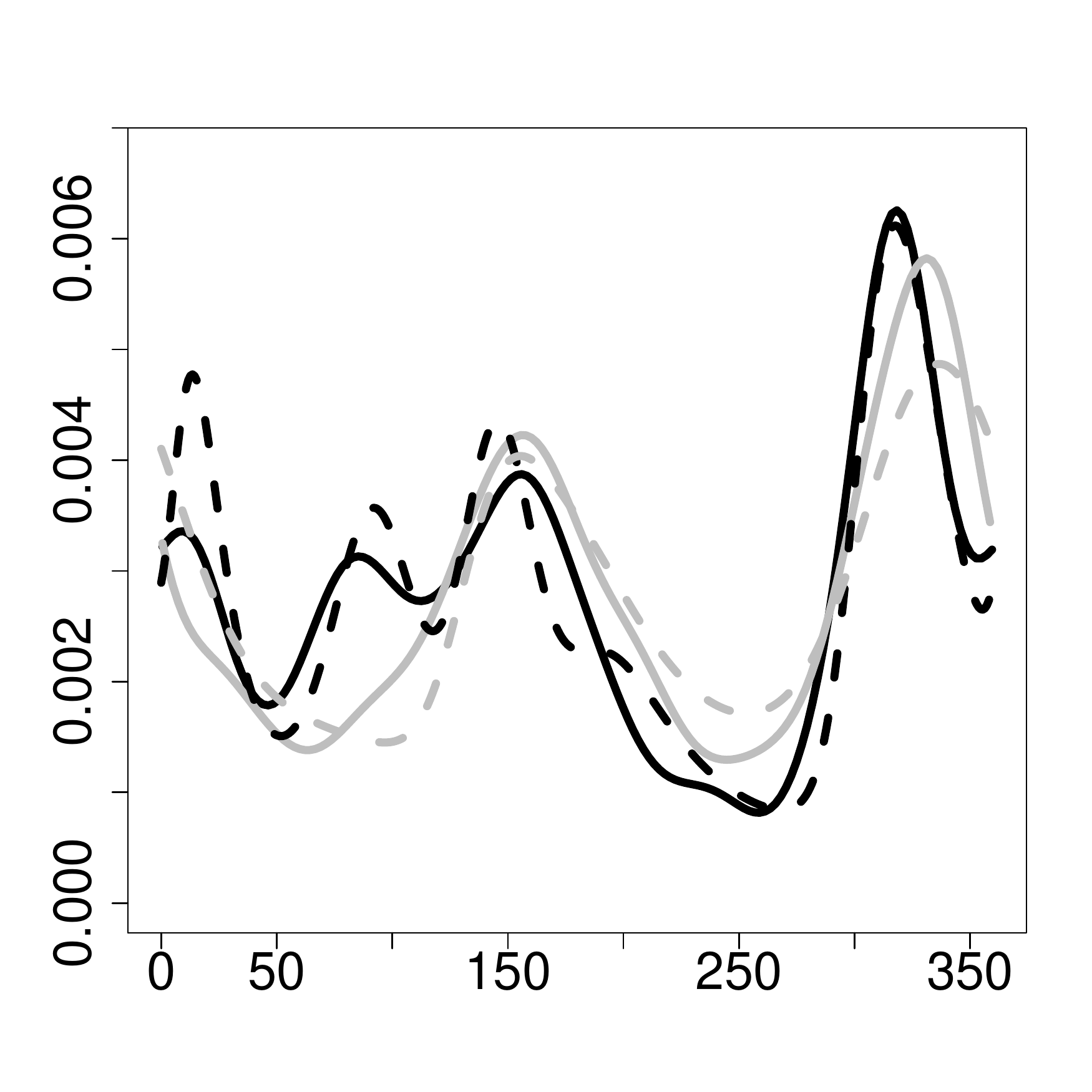}}}
    {\subfloat[WIND SPEED]{\includegraphics[trim= {0.7cm 1.cm 0.7cm 1.2cm},clip,scale=0.30]{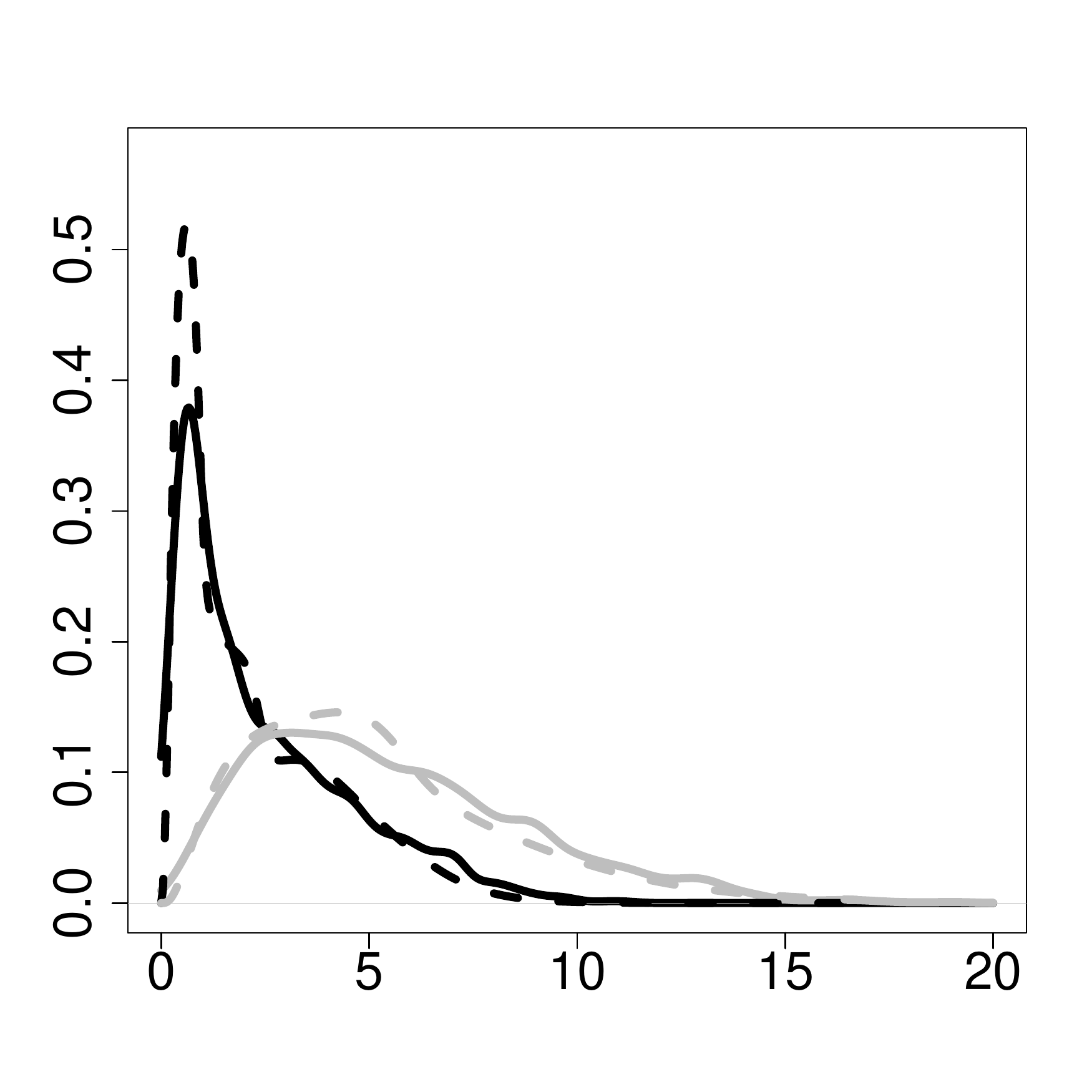}}}
    \caption{Marginal distributions of wind direction (first column) and speed (second column) in the four seasons of 2014. Black lines represent ground-observed wind speed and
        direction, grey lines are WRF-simulated. Solid lines are smooth approximations of the empirical distribution, dashed lines are sHDP-HMM-predicted distributions.}
    \label{fig:dens}
\end{figure}

The overall performance of the sHDP-HMM is described in Table \ref{tab:APE_MSE} and Figure \ref{fig:dens}. As measures of model fit we report
the average prediction error (APE) and the mean squared error (MSE) between observed and sHDP-HMM posterior predicted values, respectively
for circular and linear variables (Table \ref{tab:APE_MSE}). The APE \citep{Jona2013} is defined as the average circular distance between
observed and predicted values, where the circular distance between angles $\alpha$ and $\beta$ is given by $d(\alpha, \beta) =
1-\cos(\alpha-\beta)$. While APE takes values in [0,2), it is well known that MSE does not have a finite range. Then, for comparison, the
same indices were calculated with posterior predicted values from a minimal JPSN model for time-independent data, i.e. an i.i.d. case (in
brackets in Table \ref{tab:APE_MSE}).
%
Inspection of Table \ref{tab:APE_MSE}  shows that the sHDP-HMM time structure clearly improves the fit, measured in terms of APE and MSE,
for all variables in the four seasons. It also shows that the sHDP-HMM fits systematically better to ground-observed rather than to
WRF-simulated data. This is probably due to the known tendency of WRF to overemphasize wind peaks that are smoothed by the sHDP-HMM. As
concerns seasonal differences, it is quite clear that smoother regimes corresponding to calmer summer wind conditions favor a better fit of the
sHDP-HMM.
In Figure \ref{fig:dens} empirical and sHDP-HMM-estimated marginal distributions (solid and dashed lines) substantially agree, thus the
sHDP-HMM provides an overall reliable representation of both ground-observed and WRF-simulated wind data. Concerning the quality of WRF
forecasts, Figure \ref{fig:dens} shows that the bimodality of wind direction with peaks around the NW and SE quadrants is well reproduced
for the four seasons, together with the strong asimmetry of wind speed. WRF-simulated wind speed clearly overestimates ground recordings on
average and shows higher variability.

\begin{table}
    \centering
    \begin{tabular}{c|cccc}
        \hline\hline
        & APE$_g$ & APE$_s$ & MSE$_g$ & MSE$_s$ \\
        \hline
        WINTER &  0.515  &  0.572  &  3.835  &  13.934 \\
        & (0.969) & (0.977) & (9.109) & (30.592)\\
        \hline
        SPRING &  0.509  &  0.554  &  2.475  &   9.057 \\
        & (0.981) & (0.959) & (7.776) & (15.195)\\
        \hline
        SUMMER &  0.458  &  0.585  &  1.941  &   6.507 \\
        & (0.991) & (0.960) & (5.003) & (13.007)\\
        \hline
        AUTUMN &  0.330  &  0.667  &   3.650  &  14.323 \\
        & (0.983) & (0.995) & (11.571) & (23.998)\\
        \hline
        \hline
    \end{tabular}
    \caption{{ Values of APE and MSE for ground-observed ($g$) and WRF-simulated ($s$) data in the four seasons. In brackets the values obtained for the same indices
            by a JPSN model for time-independent data without wind regimes.}} \label{tab:APE_MSE}
\end{table}
%
%

In Table \ref{tab:par} we report the Bayesian MC estimates of ground-observed ($g$) and WRF-simulated ($s$) wind direction circular means
\eqref{eq:cmean} and concentrations \eqref{eq:cconc} and wind speed means and variances \eqref{eq:lmeanvar} for the five sHDP-HMM states
and the four seasons (Bayesian credible intervals are available in the supplementary material). Table \ref{tab:par} allows us to investigate
the main features of the detected latent homogeneous states, corresponding to winds blowing with increasing ground-observed mean wind
speed. On average, winds with higher speed show smaller variability for circular variables, i.e. stronger winds have more focused
directions. The comparison between estimated concentrations for ground and simulated data shows a generally higher variability of WRF wind
directions with respect to observed ones. Concerning forecast verification, a conservative tendency of WRF to overestimate wind speed for
all states and seasons is seen, with some difficulty in reproducing the ordering of ground-observed mean wind speed due to the
substantially higher variability. Some concern due to higher wind speed forecasts is addressed to winds blowing from SW-W (winter state 3,
spring and summer state 2). This strong positive bias may plausibly be attributed to the extreme proximity of the San Vito ground
monitoring station to the Ionian sea coast line, in the absence of the drifting effects of any obstacle to winds blowing from SW-W and
coming straight from the sea. In general, Table \ref{tab:par} shows a good agreement of ground-observed and WRF-simulated circular means,
though WRF seems to have some troubles in forecasting wind directions with low to intermediate speed (winter, spring and autumn state 1,
summer state 3). In other words, WRF always succeeds in detecting winds from the NW and SE quadrants, unless they blow with weak intensity,
as in summer.

\begin{table}
    \centering
    \begin{tabular}{cccccc}
        \hline \hline
        WINTER & 1&2&3&4&5 \\
        \hline
        $\alpha_{g,k}$ & 89.648&156.339 & 207.042&144.058 & 330.557\\
        $\alpha_{s,k}$ & 319.242& 167.336& 218.275&153.48 & 336.098\\
        $\zeta_{g,k}$ &0.548 & 0.358& 0.395&0.054 & 0.198\\
        $\zeta_{s,k}$ &0.671 &0.362 &  0.309&0.069 & 0.303\\
        $\tilde{\mu}_{g,k}$ &  0.697&1.988 &2.122  & { 4.028} & { 4.049} \\
        $\tilde{\mu}_{s,k}$ & 3.113 &3.976 &9.681 &9.922  & 6.281\\
        $\tilde{\sigma}^2_{g,k}$ & 0.147 & 1.552& 1.735 & 1.868 &  4.348\\
        $\tilde{\sigma}^2_{s,k}$ & 3.031 & 3.724 & 8.451 &14.866 & 8.444\\
        \hline \hline
        SPRING& 1&2&3&4&5 \\
        \hline
        $\alpha_{g,k}$ & 107.067&279.180 &180.703 &335.475 &313.651\\
        $\alpha_{s,k}$ &348.755  &291.028 & 169.685&347.246 &319.578\\
        $\zeta_{g,k}$ & 0.433&0.184 & 0.355& 0.373&0.065\\
        $\zeta_{s,k}$ & 0.831& 0.08& 0.235&0.626 &0.109\\
        $\tilde{\mu}_{g,k}$ & 0.722 &1.811 & 2.439& 3.277 &5.644\\
        $\tilde{\mu}_{s,k}$ & 2.709 & 5.366& 4.779&  3.816&8.247\\
        $\tilde{\sigma}^2_{g,k}$ &  0.153 & 0.843&0.842 &  1.817&4.706\\
        $\tilde{\sigma}^2_{s,k}$ & 2.568 & 4.897& 6.743& 2.857 &6.189\\
        \hline \hline
        SUMMER& 1&2&3&4&5 \\
        \hline
        $\alpha_{g,k}$ &84.871 & 296.234& 307.542& 173.961&312.629\\
        $\alpha_{s,k}$ &82.766 &261.101 & 176.688&168.684 &322.396\\
        $\zeta_{g,k}$ &0.445 &0.885 &0.305 & 0.072&0.086\\
        $\zeta_{s,k}$ &0.836 & 0.308& 0.695&0.126 &0.173\\
        $\tilde{\mu}_{g,k}$ & 0.700 &2.033 &2.337 & 2.573 &3.795\\
        $\tilde{\mu}_{s,k}$ & 2.493  & 7.368 &3.065 &4.562  &5.927\\
        $\tilde{\sigma}^2_{g,k}$ & 0.136 &1.605 &1.221 & 0.628 &1.981\\
        $\tilde{\sigma}^2_{s,k}$ & 1.524 &5.134 &2.677 & 4.129 & 4.098\\

        \hline \hline
        AUTUMN& 1&2&3&4&5 \\
        \hline
        $\alpha_{g,k}$ & 93.647& 223.014&28.571 &145.887 &325.320\\
        $\alpha_{s,k}$ &300.908 &207.719 & 55.458& 156.076&339.381\\
        $\zeta_{g,k}$ &0.120 &0.513 & 0.129&  0.044&0.061\\
        $\zeta_{s,k}$ &0.771 &0.678 & 0.589&0.082 &0.226\\
        $\tilde{\mu}_{g,k}$ & 0.627 &1.253 & 1.597&3.274  &4.813\\
        $\tilde{\mu}_{s,k}$ & 3.671 &3.653 & 4.525& 8.737 &7.672\\
        $\tilde{\sigma}^2_{g,k}$ & 0.073  & 0.592& 2.638 &  1.846 &4.273\\
        $\tilde{\sigma}^2_{s,k}$ & 3.191 & 4.789& 9.762& 11.976 &9.376\\

        \hline \hline
    \end{tabular}
    \caption{Estimates of circular means ($\alpha$) and concentrations ($\zeta$) of wind direction and means ($\tilde{\mu}$) and variances ($\tilde{\sigma}^2$) of wind speed for ground-observed ($g$) and WRF-simulated ($s$) data in the five sHDP-HMM states and four seasons. Angles are expressed in degrees and linear variables in m/s.} \label{tab:par}
\end{table}

\begin{figure}[t]
    \centering
    \captionsetup[subfigure]{labelformat=empty}
    \subfloat[]{\raisebox{+0.28in}{\rotatebox[origin=t]{90}{First regime}}}
    {\subfloat[]{\includegraphics[trim= 0 100 20 20, scale=0.16]{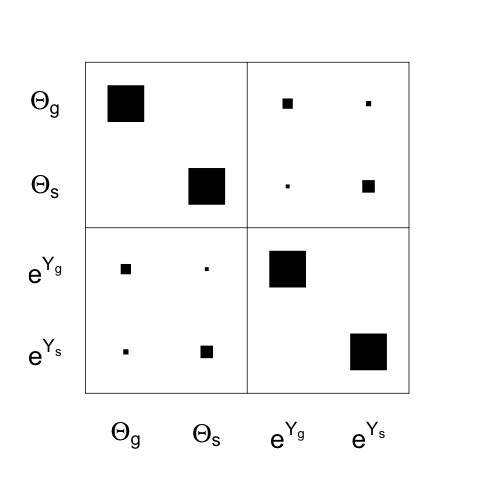}}}
    {\subfloat[]{\includegraphics[trim= 0 100 20 20, scale=0.16]{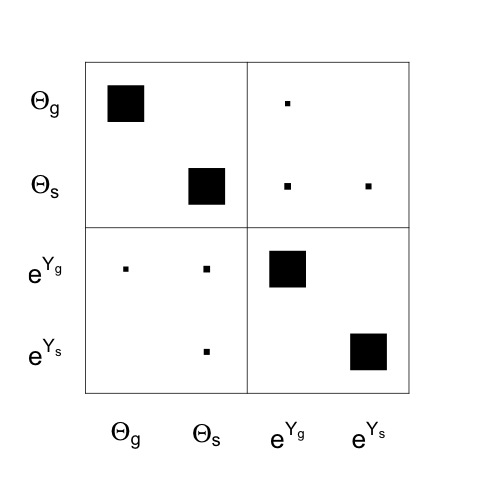}}}
    {\subfloat[]{\includegraphics[trim= 0 100 20 20, scale=0.16]{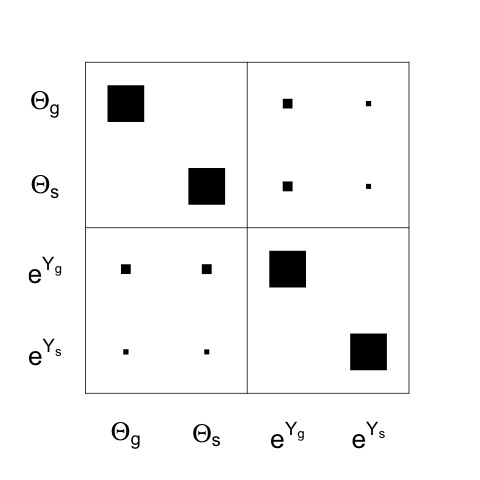}}}
    {\subfloat[]{\includegraphics[trim= 0 100 20 20, scale=0.16]{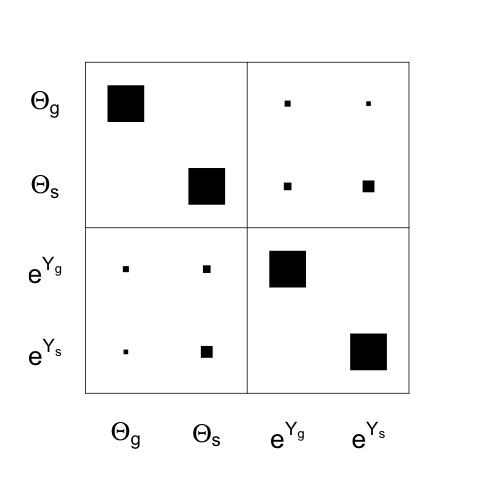}}}\\
    \subfloat[]{\raisebox{+0.26in}{\rotatebox[origin=t]{90}{Second regime}}}
    {\subfloat[]{\includegraphics[trim= 0 100 20 20, scale=0.16]{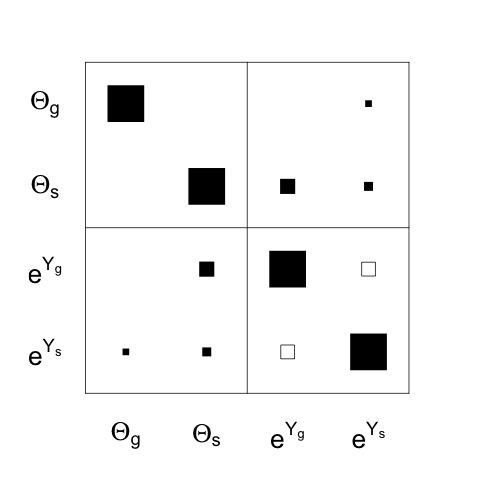}}}
    {\subfloat[]{\includegraphics[trim= 0 100 20 20, scale=0.16]{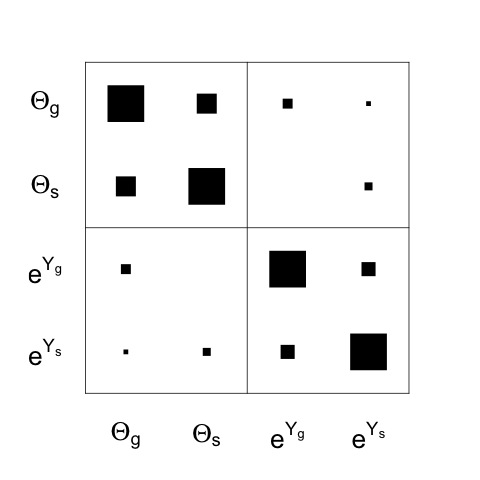}}}
    {\subfloat[]{\includegraphics[trim= 0 100 20 20, scale=0.16]{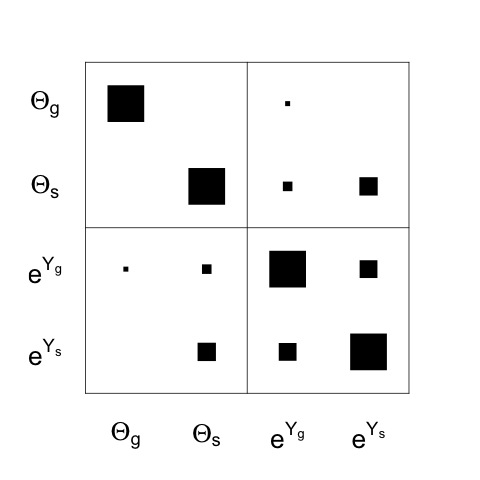}}}
    {\subfloat[]{\includegraphics[trim= 0 100 20 20, scale=0.16]{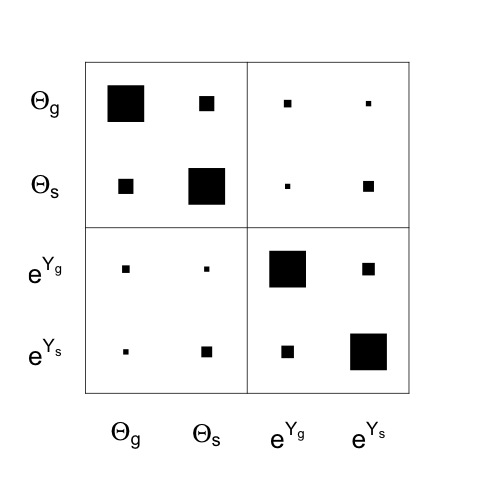}}}\\
    \subfloat[]{\raisebox{+0.25in}{\rotatebox[origin=t]{90}{Third regime}}}
    {\subfloat[]{\includegraphics[trim= 0 100 20 20, scale=0.16]{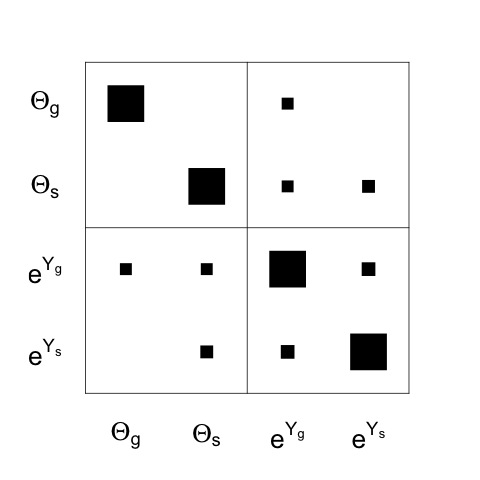}}}
    {\subfloat[]{\includegraphics[trim= 0 100 20 20, scale=0.16]{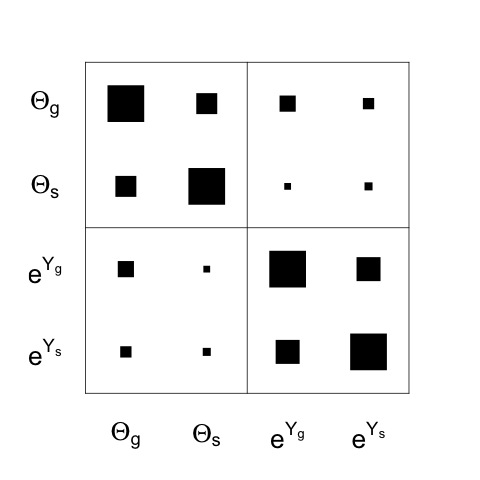}}}
    {\subfloat[]{\includegraphics[trim= 0 100 20 20, scale=0.16]{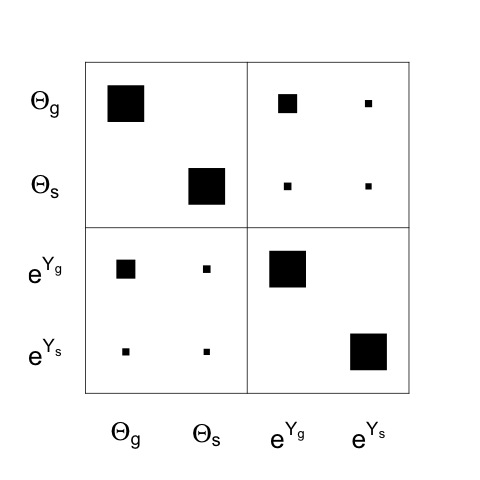}}}
    {\subfloat[]{\includegraphics[trim= 0 100 20 20, scale=0.16]{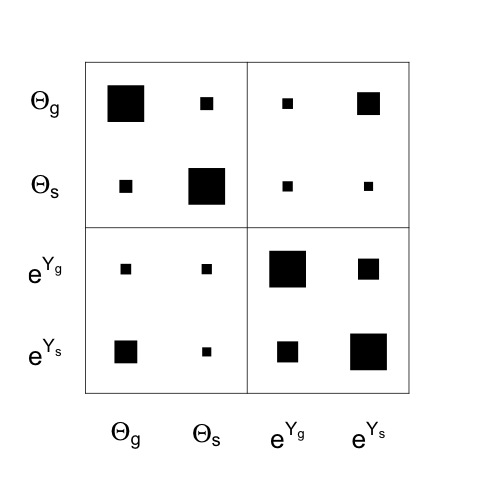}}}\\
    \subfloat[]{\raisebox{+0.25in}{\rotatebox[origin=t]{90}{Fourth regime}}}
    {\subfloat[]{\includegraphics[trim= 0 100 20 20, scale=0.16]{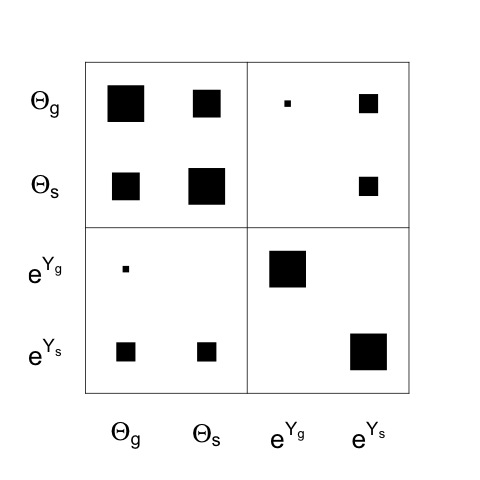}}}
    {\subfloat[]{\includegraphics[trim= 0 100 20 20, scale=0.16]{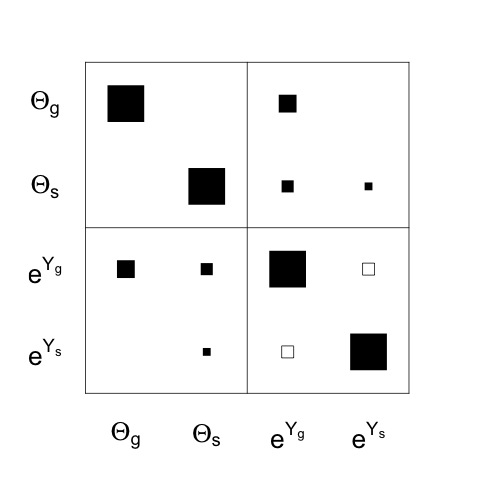}}}
    {\subfloat[]{\includegraphics[trim= 0 100 20 20, scale=0.16]{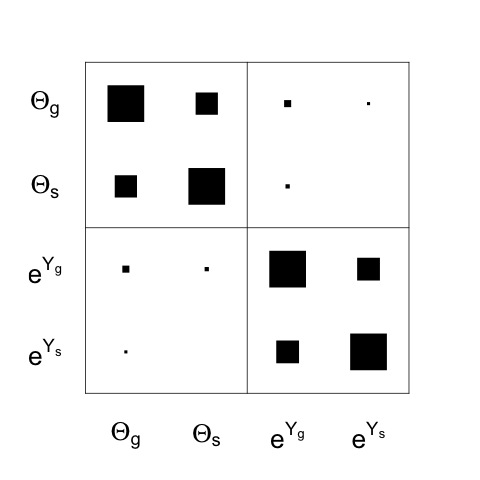}}}
    {\subfloat[]{\includegraphics[trim= 0 100 20 20, scale=0.16]{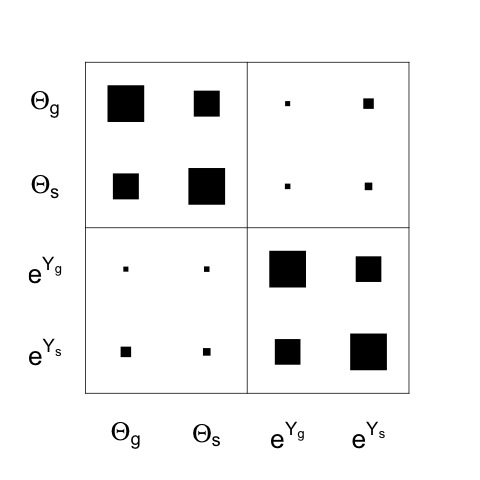}}}\\
    \subfloat[]{\raisebox{+0.46in}{\rotatebox[origin=t]{90}{Fifth regime}}}
    {\subfloat[WINTER]{\includegraphics[trim= 0 10 20 20, scale=0.16]{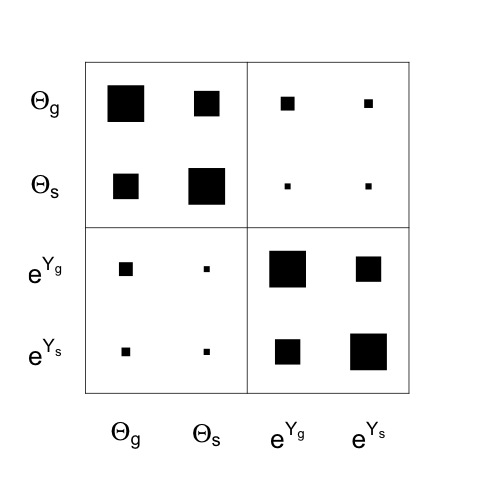}}}
    {\subfloat[SPRING]{\includegraphics[trim= 0 10 20 20, scale=0.16]{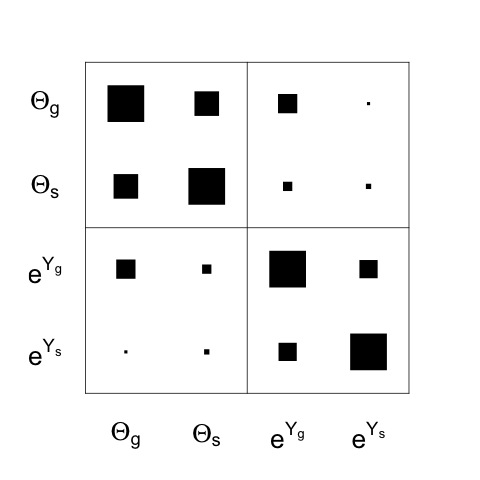}}}
    {\subfloat[SUMMER]{\includegraphics[trim= 0 10 20 20, scale=0.16]{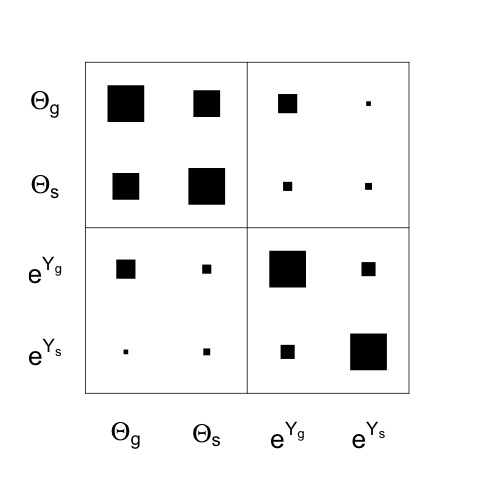}}}
    {\subfloat[AUTUMN]{\includegraphics[trim= 0 10 20 20, scale=0.16]{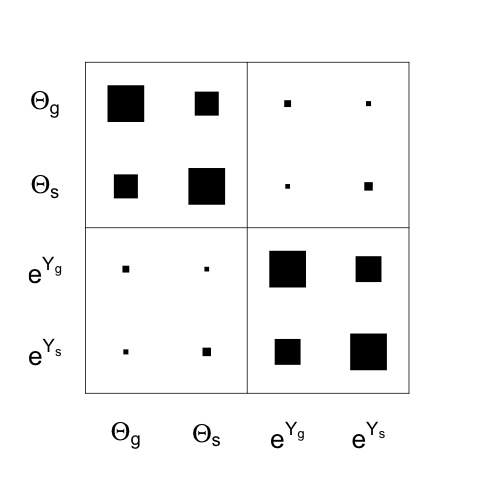}}}\\
    \caption{Dependence diagrams representing Fisher's circular-circular  correlations \eqref{eq:112}, Mardia's circular-linear  dependence measures  \eqref{eq:111} and Pearson's correlation coefficients.
        From left to right winter, spring, summer and autumn; from top to bottom the five sHDP-HMM states. Empty squares indicate negative  values as opposite to filled ones.
        Squares size is proportional to the absolute value.} \label{fig:Cor}
\end{figure}

In Figure \ref{fig:Cor}  dependences between circular and linear variables are displayed with square size proportional to the relative
association measure. Circular-circular  correlations and circular-linear  dependences \citep{fisher1996, mardia1976} are respectively computed by
\eqref{eq:112} and \eqref{eq:111}, while linear associations are measured by the Pearson's correlation coefficient. Fisher's and Pearson's
coefficients are plotted only if associated 95\% credible intervals are strictly positive or negative. Since
$\rho^2(\Theta_i,Y_j)$ in \eqref{eq:111} is null with probability 0, it was plotted according to a different criterion: notice that
$\boldsymbol{W}_i\perp Y_j\Rightarrow \Theta_i \perp Y_j$ so that if the relative elements of $\boldsymbol{\Sigma}_{wy}$ are different from
zero, then some circular-linear correlation is implied. Accordingly, posterior mean values of $\rho^2(\Theta_i,Y_j)$ are plotted in Figure
\ref{fig:Cor} only if at least one of the 95\% credible intervals of the associated components of $\boldsymbol{\Sigma}_{wy}$ does not
contain zero.
Notice the high number of boxes with negative, null or weak correlation in states 1, 2 and 3, while states 4 and 5, corresponding to higher
observed wind speed, are more likely to show positive correlations between ground-observed and WRF-simulated wind components. In
particular, no meaningful correlations between forecasts and observations are estimated for state 1. Circular-linear associations are
overall weaker, but still show the same tendency to increase with ground-observed wind speed.

\section{Concluding remarks}

In the previous Section, the sHDP-HMM was used to reproduce ground data, WRF simulations and the relationship between the two.
Here we do not focus on the calibration of the WRF system: rather, the description of the features that characterize clusters of observed
and forecasted winds is the fundamental output of the proposed method in terms of forecast verification. Among these features are not only
the distribution summaries for both observed and simulated data, but also correlations between variables and transition probabilities of
moving from one sHDP-HMM component to another, informing on the time evolution of homogeneous states (Table 2 in the supplementary
material). The distributions-oriented approach allows us to deal with forecast verification in a fully comprehensive model estimation setting.
In addition, many uncertainty features of the process and its components are quantified in the Bayesian estimation framework (as can be
appreciated by the posterior credibility intervals of model parameters, fully reported in the supplementary material).
In this respect, notice that the remarkably different variability of WRF with respect to ground-observed wind speed records, reproduced by the
sHDP-HMM as reported in Table \ref{tab:par}, may be due the peculiar position of the validation point, located in a complex
residential/industrial area characterized by the extreme proximity to the coast line of the Ionian Sea. At
point locations considerably affected by local features, as the San Vito station, the spatial discretization required for the numerical
solution of the atmospheric equations affects the variability of WRF-simulated wind records given by averages over volumes with homogeneous
properties \citep{jimenez2010surface}. Within these simulation volumes, the smoothing of surface physical properties (e.g. orography)
implied by the spatial discretization has also a strong influence on wind simulations, producing higher smoothness and variability. Besides
showing the ability of the sHDP-HMM in reproducing both ground-observed and WRF-simulated wind data, our proposal allows us to investigate some
peculiar characteristics of the WRF system performance at a specific validation point. Within an estimation framework that allows full
control of process and model uncertainties, it highlights features not as precisely derived within the traditional measurements-based
forecast verification framework.
This distributions-oriented proposal can then be regarded as a way to check the reliability of the WRF simulations for the specific
purposes addressed in Section \ref{s:intro}. In fact the ability to simulate focused directions for strong winds makes the application of
WRF (with the specific settings reported in Table \ref{tab:tab1}) especially suitable for forecasting strong wind events as defined by the
Regional Air Quality Plan with the aim to reduce industrial emission during such events.

Among the possible developments of the present proposal is the investigation of the possibility to model the seasonality in the observed
process. For this purpose, the regime switching mechanism might be driven by a non-homogeneous Markov chain whose transition probabilities
are periodic functions. { In  all the work, the assumption that the amount of null wind speed recordings and their dependence on the data generation process are negligible is justified from a physical point of view. In a more realistic situation, zeros could be considered as missing values within a latent variable approach and predicted along with unknown parameters during model fitting.}


\section*{Acknowledgement}
This work is partially developed under the PRIN2015 supported-project ``Environmental processes and human activities: capturing their interactions via statistical methods (EPHASTAT)'' funded by MIUR (Italian Ministry of Education, University and Scientific Research).

\bibliographystyle{natbib}
\bibliography{all}

\end{document}